\documentclass[preprint,preprintnumbers,amsmath,floatfix]{revtex4}
\usepackage{amsfonts}    
\usepackage{amssymb}
\usepackage{latexsym}
\usepackage{eepic}
\usepackage{graphicx}
\usepackage{subfig}
\usepackage{subfig}
\usepackage{multirow}
\usepackage{rotating,booktabs} 
\usepackage{pdflscape}
\usepackage[active]{srcltx}
\usepackage[colorlinks=true]{hyperref}
\usepackage{caption}
\usepackage{color}
\newcommand{\br}{\rule[0.5cm]{0.001cm}{0.0cm}}  

\newcommand{\re}[1]{(\ref{#1})}

\newcommand{\al}{\alpha}
\newcommand{\pa}{\partial}

\newcommand{\La}{\Lambda}
\newcommand{\si}{\sigma}
\newcommand{\la}{\lambda}

\newcommand{\ga}{\gamma}

\newcommand{\de}{\delta}

\newcommand{\De}{\Delta}

\newcommand{\tha}{\theta}

\newcommand{\rar}{\rightarrow}
\newcommand{\lrar}{\leftrightarrow}
\newcommand{\non}{\nonumber}
\newcommand{\m}{\,\,}
\begin{document}

\title{The H$_2^+$ molecular ion: low-lying states}

\author{Horacio~Olivares-Pil\'on} \email{horop@nucleares.unam.mx}
\affiliation{Departamento de F\'isica, Universidad Aut\'onoma
  Metropolitana-Iztapalapa, Apartado Postal 55-534, 09340 M\'exico,
  D.F., Mexico}

\author{Alexander~V.~Turbiner}
\email{turbiner@nucleares.unam.mx}
\affiliation{Instituto de Ciencias Nucleares, Universidad Nacional
Aut\'onoma de M\'exico, Apartado Postal 70-543, 04510 M\'exico,
D.F., Mexico{}}


\vskip 1cm

\begin{abstract}
Matching for a wavefunction the WKB expansion at large distances and Taylor
expansion at small distances leads to a compact, few-parametric uniform
approximation found in {\it J. Phys. B44, 101002 (2011)}.
The ten low-lying eigenstates of H$_2^+$ of the quantum numbers
$(n,m,\La,\pm)$\, with $n=m=0$ at $\La=0,1,2$, with $n=1$, $m=0$ and $n=0$,
$m=1$ at $\La=0$ of both parities are explored for all interproton distances $R$.
For all these states this approximation provides the relative accuracy $\lesssim 10^{-5}$
(not less than 5 s.d.) locally, for any real coordinate $x$ in eigenfunctions,
when for total energy $E(R)$ it gives 10-11 s.d. for $R \in [0,50]$~a.u.
Corrections to the approximation are evaluated in the specially-designed,
convergent perturbation theory. Separation constants are found with not
less than 8 s.d. The oscillator strength for the electric dipole transitions
$E1$ is calculated with not less than 6~s.d.
A dramatic dip in the $E1$ oscillator strength $f_{1s\si_g-3p\si_u}$
at $R \sim R_{eq}$ is observed. The magnetic dipole and electric quadrupole
transitions are calculated for the first time with not less than 6~s.d.
in oscillator strength. For two lowest states $(0,0,0,\pm)$ (or,
equivalently, $1s\si_g$ and $2p\si_u$ states) the potential curves are checked and
confirmed in the Lagrange mesh method within 12~s.d. Based on them
the Energy Gap between $1s\si_g$ and $2p\si_u$ potential
curves is approximated with modified Pade $R e^{-R} [Pade(8/7)](R)$
with not less than 4-5 figures at $R \in [0, 40]$\,a.u.
Sum of potential curves $E_{1s\si_g} + E_{2p\si_u}$ is approximated
by Pade $1/R [Pade(5/8)](R)$ in $R \in [0, 40]$\,a.u. with not less
than {3-4} figures.
\end{abstract}


\maketitle

\centerline{\bf INTRODUCTION}

\vskip 1cm

The H$_2^+$ molecular ion is the simplest molecular system which
exists in Nature. It plays a fundamental role in atomic-molecular
physics, in laser and plasma physics being also a traditional example
of two-center Coulomb system of two heavy Coulomb charges $Z$ and electron,
$(Z, Z, e)$ in Quantum Mechanics (see e.g. \cite{LL}). Due to the fact that
the proton is much heavier than electron the problem is usually explored in
the static approximation - the Bohr-Oppenheimer approximation of the zero
order - where the protons are assumed to be infinitely heavy. In general,
the projection of the angular momentum to the molecular axis (the line
connecting the proton positions) $L_{\phi}$ is the integral,
$[L_{\phi}, {\cal H}]=0$, where ${\cal H}$ is the Hamiltonian. Thus,
the angular variable $\phi$ can be separated out. Hence, the problem is
reduced to two-dimensional, which admits itself the separation of variables
in elliptic coordinates.
It reflects the outstanding property of the general two-center Coulomb problem
$(Z_1, Z_2, e)$ of the complete separation of variables in prolate
ellipsoidal coordinates.

General two-center Coulomb problem $(Z, Z, e)$ can not be solved exactly, but
approximately only. Thus, we need to introduce a definition of {\it solvability}
of the spectral problem, of the corresponding Schr\"odinger equation:
for all (or some) eigenfunctions $\Psi(x)$ we have to be able to find
constructively an uniform approximation $\Psi_{app}(x)$ such that
\begin{equation}
\label{delta}
  \left| \frac{\Psi(x) - \Psi_{app}(x)}{\Psi_{app}(x)} \right| \lesssim 10^{-\de} \ ,
\end{equation}
in the coordinate space, while in vicinity of the nodal surface, $\Psi_{app}(x)=0$,
the absolute deviation
\begin{equation}
\label{delta-n}
  \left| {\Psi(x) - \Psi_{app}(x)} \right| \lesssim 10^{-\de} \ .
\end{equation}
The parameter $\de > 0$ characterizes a number of significant digits
(s.d.) in wavefunction at real $x$, which the approximation reproduces
exactly. It implies that any observable, any matrix element can be
found with accuracy not less than $\de$. In principle, in the case of
non-relativistic QED in the Born-Oppenheimer (static) approximation we
think that $\de \sim {4 - 5}$ is sufficient to get physically-relevant
results: the corrections due to finite proton mass, its form factor,
relativistic effects of different types are small;
in particular, for energies of states they should contribute to significant
digit 4,5,6 etc. Our aim is to solve the problem of H$_2^+$ molecular ion in
non-relativistic QED approximation by constructing maximally-simple, compact,
locally-accurate approximations for the ten low-lying eigenfunctions.

The goal of this paper is to extend and profound the analysis in \cite{Turbiner:2011} for the
states $1s\si_{g}$ and $2p\si_{u}$ and explore the eight more low lying states of the H$^+_2$
molecular ion. In order to check accuracy of obtained approximations a convergent perturbation
theory (PT) used in \cite{Turbiner:2011} is extended for the case of excited states.
This PT allows us to evaluate a local deviation of the approximation from the exact eigenfunction. Eventually, we calculate systematically separation constants and the oscillator strength for the
electric $E1$ dipole and $E2$ quadrupole, and magnetic $B1$ dipole transitions.

It is worth mentioning that a study of the wavefunctions of the H$^+_2$ molecular ion in a form of expansion in some basis was initiated a long ago by Hylleraas \cite{Hylleraas:1931}, and it was successfully realized in the remarkable paper \cite{Bates:1953} (see also \cite{Montgomery:1977, Bishop:1978}). Since old times there were made many attempts to find bases leading to fast
convergence. At present, the basis of pure exponential functions seems the most fast convergent
(see e.g. \cite{Korobov:2000} and references therein). Note that following the analysis of classical mechanics of the H$_2^+$ system and its subsequent semiclassical quantization it was attempted a long ago to build some compact uniform approximations of wavefunctions of low lying electronic states \cite{Strand:1979}. Local accuracies of these approximations are unclear whilst eigenvalues
are found with a few significant digits. We are unaware about further attempts in this direction
except for our previous paper \cite{Turbiner:2011}. Note that a similar idea to construct
compact uniform approximations of the lowest eigenfunctions was successfully realized for quartic
anharmonic oscillator \cite{Turbiner:2005} and double-well potential \cite{Turbiner:2010}.

Throughout the paper the Rydberg is used as the energy unit while for the other quantities standard
atomic units are used $\hbar=m_e=e=1$.

\section{Generalities}

\begin{figure}[!thb]
\includegraphics[scale=0.5]{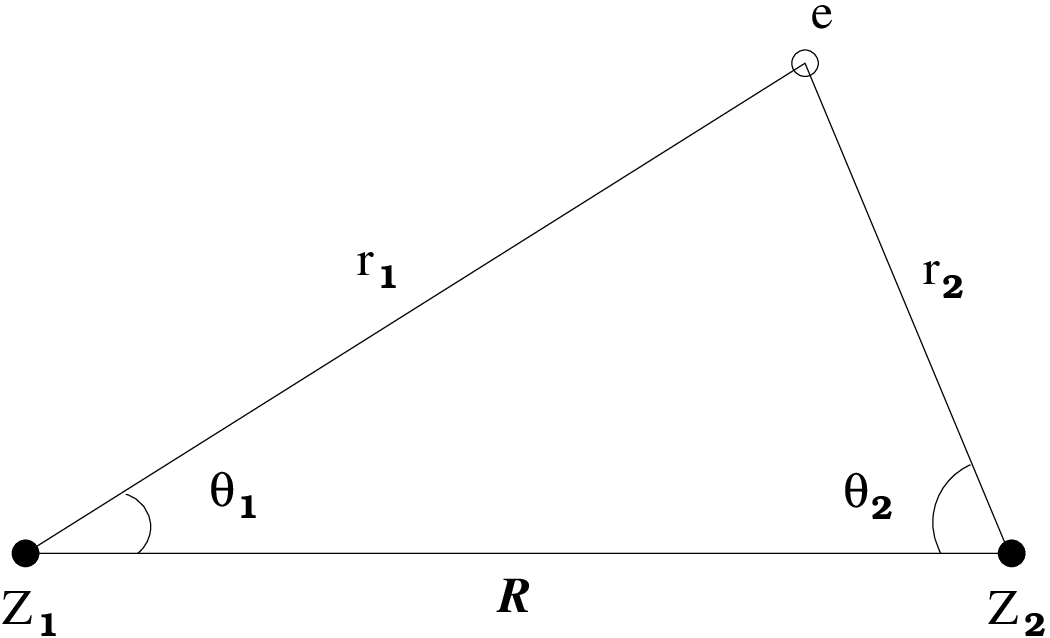}
\caption{Geometrical settings for $(Z_1,Z_2,e)$.}
\label{geometry}
\end{figure}

The Schr\"odinger equation, which describes the electron in the field of two fixed centers of the charges $Z_1, Z_2$ at the distance $R$, is of the form
\begin{equation}
\label{Sch}
    \left(-\De - \frac{2 Z_1}{r_1}- \frac{2 Z_2}{r_2}\right)\Psi \ =\ E' \Psi\ ,\
    \Psi \in L^2 ({\bf R^3})\ ,
\end{equation}
where $E'=(E - \frac{2 Z_1 Z_2 }{R})$ and the total energy $E$ are in Rydbergs, $r_{1,2}$ are the distances from electron to first (second) center, respectively, see Fig.~\ref{geometry}. All distances are in a.u. From physical point of view, we study the motion of electron in the field of two Coulomb wells situated on the distance $R$. Hence, if $Z_1=Z_2$ the wells become identical - any eigenstate should be characterized by a definite parity with respect to permutation of wells. Furthermore, at $R \rar \infty$, when the barrier gets large and tunneling becomes exponentially-small, the phenomenon of pairing should occur: the spectra of positive parity states is almost degenerate with the spectra of negative parity states. For each pair the energy gap should be exponentially-small, $\sim e^{-R}$.

Following \cite{LL} let us introduce the dimensionless 2D elliptic coordinates and azimuthal angle  $\varphi$ with respect to the molecular axis
\footnote{From $3D$ point of view they are {\it prolate spheroidal}.}:
\begin{equation}
\label{ell}
  \xi = \frac{r_1+r_2}{R}\ ,\quad \eta = \frac{r_2-r_1}{R}\ ,\quad 1 \leq \xi \leq \infty\ ,\quad -1 \leq \eta \leq 1\ .
\end{equation}
In these coordinates the Coulomb singularities are situated at
\[
 \xi=1\quad ,\quad \eta = \pm 1\ ,
\]
being at the boundaries of the configuration space.
The Jacobian is $\propto (\xi^2-\eta^2)$. The equation (\ref{Sch}) admits separation of variables in (\ref{ell}). Since the projection of the angular momentum to the molecular axis $\hat L_{\phi}$ commutes with the Hamiltonian
\footnote{Due to complete separation of variables one more integral in a form of the second order polynomial in momentum exists \cite{Erikson:1949}, it is closely related to Runge-Lenz vector \cite{Coulson:1967} and commutes with $\hat L_{\phi}$; hence, the H$_2^+$ ion in adiabatic (Born-Oppenheimer) approximation is completely-integrable system.}
the eigenstate has a definite magnetic quantum number $\La$. If $Z_1=Z_2$ the Hamiltonian is permutationally-symmetric $r_1 \lrar r_2$, or, equivalently, $\eta \rar -\eta$, hence, any eigenfunction is of a definite parity ($\pm$). As a result, it can be represented in a form
\begin{equation}
\label{psi}
   \Psi \ =\ X(\xi) (\xi^2-1)^{\La/2} Y(\eta) (1-\eta^2)^{\La/2} e^{ \pm i \La \phi}\ ,\ \La=0,1,2,\ldots
\end{equation}
where $Y$ is of definite parity, $Y(\eta)=\eta^K Z(\eta^2), K=0,1$. Following this analysis we introduce the notation for a state as $(n,m,\La,\pm)$ where $n,m=0,1,\ldots$ are the quantum numbers in $\xi$ and $\eta$ coordinates, respectively, they have a meaning of number of nodes in $\xi$ and $\eta$, $\La$ is a magnetic quantum number, while $\pm=(-1)^K$ is parity. It is easy to check that the ground state with the lowest total energy is $(0,0,0,+)$.

The factors $(\xi^2-1)^{\La/2}$ and $(1-\eta^2)^{\La/2}$ are introduced to (\ref{psi}) to take into account a singular behavior of the eigenfunction near Coulomb singularities in accordance to the boundary conditions. After substitution of the representation (\ref{psi}) into (\ref{Sch}) we arrive at the equations for $X(\xi)$ and $Y(\eta)$,
\begin{equation}
\label{X_L}
   L_{\xi}\,X \equiv [\pa_{\xi} (\xi^2-1) \pa_{\xi}]\, X + 2 \La \xi \pa_{\xi}\,X - (p^2\xi^2 - 2R \xi)\, X
   \ =\ -A\,X\ , \ X \in L^2(\xi \in [1,\infty))\ ,
\end{equation}
\begin{equation}
\label{Y_L}
    L_{\eta}\,Y \equiv  [\pa_{\eta} (\eta^2-1) \pa_{\eta}]\, Y + 2 \La \eta \pa_{\eta}\, Y -p^2\eta^2\, Y
    \ =\ -A\,Y\ , \ Y \in L^2(\eta \in [-1,1])\ ,
\end{equation}
respectively, where following \cite{Bates:1953} we denote,
\begin{equation}
\label{p}
    p^2\ =\ - \frac{E'R^2}{4}\ ,
\end{equation}
and $A$ is a separation constant. Equations (\ref{X_L}), (\ref{Y_L}) define a bispectral problem with $E,A$ as spectral parameters for any given $R$.

It is interesting that the operators in lhs of the equations (\ref{X_L}), (\ref{Y_L}) are of the Lie-algebraic nature. After the gauge rotation
\[
 {\ell}_{\xi} \equiv e^{p\xi}\, L_{\xi}\, e^{-p\xi} = (\xi^2-1)\pa_{\xi}^2 - 2 [p \xi^2 - (\La+1)\xi +p]\pa_{\xi}-2[p(\La+1)-R]\xi - p^2\ ,
\]
the operator can be rewritten in terms of the $sl(2)$-Lie algebra generators
\begin{equation}
\label{sl2}
    J_-\ =\ {\pa_{\xi}}\quad ,\quad J_0\ =\ {\xi} {\pa_{\xi}} - \frac{\nu}{2}\ ,\
    J_+\ =\ {\xi} ({\xi} {\pa_{\xi}} - \nu)\ ,
\end{equation}
see e.g. \cite{Turbiner:1988}, with
\[
      \nu_{\xi} \ =\ -\La - 1 + \frac{R}{p}\ .
\]
In a similar way after the gauge rotation
\[
 {\ell}_{\eta} \equiv e^{-p\eta}\, L_{\eta}\, e^{p\eta} = (\eta^2-1)\pa_{\eta}^2 + 2 [p \eta^2 + (\La+1)\eta +p]\pa_{\eta}+2p(\La+1)\eta - p^2\ ,
\]
the operator can be rewritten in terms of the $sl(2)$-Lie algebra generators (\ref{sl2})
with
\[
      \nu_{\eta} \ =\ -\La - 1 \ .
\]
Hence, the hidden algebra of H$_2^+$ molecular ion after separation of the angular variable $\phi$ in coordinates $\xi, \eta$ is $sl(2)\oplus sl(2)$ \cite{Turbiner:2011}.
The spin of the representation is $-\La-1+\frac{R}{p}$ and $-\La-1$, respectively. For non-physical, integer, non-negative values of $-(\La+1)$ and integer ratio $\frac{R}{p}$, each algebra $sl(2)$
appears in the finite-dimensional representation realized in action on
polynomials in $\xi, \eta$, respectively. It is worth noting that the operator $L_{\eta}$ (\ref{Y_L})
is $Z_2$-invariant: $\eta \rar -\eta$. Thus, it can be gauge-rotated with $\eta^K, K=0,1$ and then a new variable $\eta_2=\eta^2$ can be introduced,
\[
    \tilde {\ell} \equiv \left. \eta^{-K}\,{\ell}_{\eta}\,\eta^K\,\right |_{\eta_2=\eta^2}\ =\ 4\eta_2(\eta_2-1) \pa_{\eta_2}^2 + 2[(2\La+3+2K)\eta_2 -(2K+1)]\pa_{\eta_2} + 2K(\La +1)-p^2\eta_2\ ,
\]
it leads to another convenient representation of the Eq.(\ref{Y_L}).

Square-integrability of the function $\Psi$ (\ref{psi}) implies a non-singular behavior of $X$ at $\xi \rar 1$ and decay at $\xi \rar \infty$ as well as non-singular behavior of $Y$ at $\eta \rar \pm 1$. Such a non-singular solution $X$ can be continued from the interval $[1, +\infty)$ to the whole line $(-\infty, +\infty)$. It implies searching a solution of the spectral problem (\ref{X_L}) which grows at $\xi \rar -\infty$, decays at $\xi \rar +\infty$ being a constant at $\xi=1$. A non-singular solution $Y(\eta)$ at $\eta=\pm 1$ can be unambiguously continued in $\eta$ beyond the interval $[-1,1]$ to $(-\infty, +\infty)$, it corresponds to growing (non-decaying) behavior of $Y(\eta)$ at $|\eta| \rar \infty$.

The equation (\ref{X_L}) formally coincides with equation (\ref{Y_L}) at $R=0$ (united atom limit). It is evident that if a domain for (\ref{Y_L}) is extended to $[1,\infty)$ it has no $L^2$ solutions since there is no degeneracy at any $R$ with $R=0$. Hence, at $E,A$ the solution found solving the equation (\ref{X_L}) should be non-normalizable. Since at $R=0$ the problem becomes one-center Coulomb problem and can be solved exactly. The above statement can be checked explicitly.
It is in agreement with large $\xi$-behavior of the celebrated Guillemin-Zener function, which mimics the coherent interaction of electron with charged centers, see e.g. \cite{Turbiner:2011}. It leads to
\begin{equation}
\label{GZ}
   \Psi^{(\pm)}_{GZ}\ =\ e^{- 2\al_3 r_1 - 2\al_4 r_2} \pm e^{- 2\al_3 r_2 - 2\al_4 r_1}\
   =\ 2 e^{-(\al_3+\al_4) R \xi}
   \left[
   \begin{array}{c}
    \cosh ((\al_3-\al_4) R \eta) \\
    \sinh ((\al_3-\al_4) R \eta)
    \end{array}
    \right]
    \ ,
\end{equation}
which eventually has to describe small $R$ behavior of the H$_2^+$. It is also in agreement with large $\xi$-behavior of the celebrated Hund-Mulliken function (it mimics the incoherent interaction of electron with charged centers) for both $1s\si_g$ (of positive parity) and  $2p\si_u$ (of negative parity) states, see e.g. \cite{Turbiner:2011},
\begin{equation}
\label{HM}
    \Psi^{(\pm)}_{HM}\ =\ e^{- 2\al_2 r_1} \pm e^{- 2\al_2 r_2}\
    =\ 2 e^{-\al_2 R \xi}
    \left[
   \begin{array}{c}
    \cosh (\al_2 R \eta) \\
    \sinh (\al_2 R \eta)
    \end{array}
    \right]
    \ ,
\end{equation}
which describes large $R$ behavior.

\subsection{\it Asymptotics.} Assuming a representation $X=e^{-\varphi}$, then the WKB-type-expansion of phase $\varphi$ at $\xi \rar \infty$ can be derived,
\begin{equation}
\label{X-inf}
 \varphi\ =\ p\xi - \left( \frac{R}{p}-\La-1 \right) \log \xi + \left[\frac{A+(\frac{R}{p}-\La-1)(\frac{R}{p}+\La)}{p} - p\right]\frac{1}{2\xi}
 +\ldots\ ,
\end{equation}
while at $\xi \rar 0$,
\begin{equation}
\label{X-small}
 \varphi\ =\ -\frac{A}{2}\xi^2 - \frac{R}{3}\xi^3 + \frac{(p^2+A^2-A(2\La+3))}{12} \xi^4\
 +\ \ldots \ .
\end{equation}
Similarly to $X$ if we assume $Y = e^{-\varrho}$, then at $\eta \rar \infty$,
\begin{equation}
\label{Y-inf}
 \varrho\ =\ -p\eta + (\La+1)\log \eta - \left(\frac{A-\La(\La+1)}{p} - p\right)\frac{1}{2\eta}
 +\ldots\ ,
\end{equation}
when at $\eta \rar 0$,
\begin{equation}
\label{Y-small}
 \varrho\ =\ -\frac{A}{2}\eta^2 + \frac{(p^2+A^2-A(2\La+3))}{12} \eta^4 + \ldots\ .
\end{equation}
The important property of the expansions (\ref{X-inf}) and (\ref{Y-inf}) is that the coefficients in front of the growing terms at large distances (linear and logarithmic)
are found explicitly, since they do not depend on the separation constant $A$.

\subsection{\it Approximation}
Making interpolation between WKB-type-expansion (\ref{X-inf}) and the perturbation theory (\ref{X-small}) for $X$, (\ref{Y-inf}) and (\ref{Y-small}) for $Y$, correspondingly, and taking into account that the $Z_2-$symmetry: $\eta \rar -\eta$ of $\Psi$ is realized through use of $\cosh(\sinh)$-function (cf. (\ref{HM}) and (\ref{GZ})) we arrive at the following expression \cite{Turbiner:2011}
\[
 \Psi^{(\pm)}_{n,m,\La}\ =
\]
\begin{equation}
\label{appr}
\frac{(\xi^2-1)^{\La/2}P_n (\xi)}{(\gamma + \xi)^{1+n+\La-\frac{R}{p}}}
 e^{-\xi \frac{\al + p \xi}{\gamma + \xi}}
 \frac{(1-\eta^2)^{\La/2}Q_m(\eta^2)}{(1 + b_2 \eta^2 + b_3 \eta^4)^{\frac{1+2m+\La}{4}}}
 \left[
   \begin{array}{c}
    \cosh  \\
    \sinh
    \end{array}
 \left(\eta \frac{a_1 + p a_2 \eta^2 + p b_3 \eta^4}
 {1 + b_2\eta^2 + b_3 \eta^4}\right) \right] e^{\pm i \La \phi} \ ,
\end{equation}
for the eigenfunction of the state with the quantum numbers $(n,m,\La,\pm)$. Here $\al,\gamma$ and $a_{1,2}, b_{2,3}$ are parameters (see below), $P_n (\xi)$ and $Q_m(\eta^2)$ are some polynomials of degrees $n$ and $m$ with real coefficients with $n$ and $m$ real roots in the intervals $[1,\infty)$ and $[0,1]$, respectively. We will choose these polynomials in such a way to ensure their orthogonality to all states with lower total energies.

\section{Results}

\subsection{Ground state of positive/negative parity}

In this Section we consider briefly two lowest states - one of positive and one of negative parity, $1s \si_g\ (0,0,0,+)$ and $2p \si_u\ (0,0,0,-)$, respectively, updating the consideration which was done in
\cite{Turbiner:2011}. Corresponding approximations for above states have the form
\[
 \Psi^{(\pm)}_{0,0,0} = \frac{1}{(\gamma + \xi)^{1-\frac{R}{p}}}
 e^{-\xi \frac{\al + p \xi}{\gamma + \xi}}
 \frac{1}{(1 + b_2 \eta^2 + b_3 \eta^4)^{1/4}}
 \left[
   \begin{array}{c}
    \cosh  \\
    \sinh
    \end{array}
 \left(\eta \frac{a_1 + p a_2 \eta^2 + p b_3 \eta^4}
 {1 + b_2\eta^2 + b_3 \eta^4}\right) \right]
\]
\begin{equation}
\label{appr-0}
   \equiv X_0(\xi) Y_0^{(\pm)}(\eta) \ ,
\end{equation}
(cf.(\ref{appr})), respectively, and each of them depends on six
parameters $\al,\gamma$ and $a_{1,2}, b_{2,3}$. The easiest way to
choose these parameters is to make a variational calculation taking
(\ref{appr-0}) as a trial function for $R$ fixed and with $p$ as an
extra variational parameter. Immediate striking result of the
variational study is that for all $R \in [1,50]$~a.u. the optimal value of
the parameter $p$ coincides with the exact value of $p$ (see
(\ref{p})) with extremely high accuracy for both $1s \si_g$ and $2p
\si_u$ states. It implies a very high quality of the trial function -
the variational optimization wants to reproduce with very high
accuracy a domain where the eigenfunction is exponentially small,
hence, the domain which gives a very small contribution to the energy
functional. In Tables \ref{Ten1ssg} and \ref{Ten2psu} some selected results for
the total energy (as well as for sensitive $p$) vs $R$ of $1s \si_g$
and $2p \si_u$ states are shown as well as their comparison with ones
obtained by Montgomery \cite{Montgomery:1977} (in highly-accurate
realization of the approach proposed by Bates et al \cite{Bates:1953}),
and also with the results we obtained in the Lagrange mesh method based on
Vincke-Baye approach \cite{Baye:2006} (for details and discussion
see \cite{Baye:2015}).
For all studied values of $R$ for both $1s \si_g$ and $2p \si_u$ states our
variational energy turns out to be in agreement on the level of 10 s.d. with
these two alternative calculations.
Note that for $R \sim 30$\,a.u. the energies of $1s \si_g$ and $2p\si_u$ states
coincide in 10 s.d. according to the pairing phenomenon. Also they differ from
the ground state energy of Hydrogen atom $\sim 10^{-6}$\,a.u., this difference
reduces gradually with further growth of $R$.

The $1s\si_g$ potential curve has a ``flat" minimum at $R=R_{eq}=1.997\,193$\,a.u.
characterized by a small curvature, see Table \ref{Ten1ssg}. It defines the
equilibrium distance $R_{eq}$: numerically, the value of $R_{eq}$ agrees
with \cite{B:1970} in all seven digits. The first non-vanishing term in a Taylor
expansion of the $1s\si_g$ potential curve around the minimum at $R=R_{eq}$ is
quadratic $O((R-R_{eq})^2)$. Thus, the coefficient, proportional to the so-called
harmonic force constant $k_e$, see e.g. \cite{B:1970}, is given by the expression
\begin{equation}
     k_e\ =\ \left. \frac{d^2 E(R)}{dR^2}\right|_{R=R_{eq}}\ .
\end{equation}
Making accurate calculations around minimum of the $1s\si_g$ potential curve
one can calculate $k_e$. Not surprisingly, it coincides with value
$k_e = 0.205940876$ reported by Bishop~\cite{B:1970} in all digits.

The $2p \si_u$ potential curve, inside the accuracy of our calculations,
displaces very ``flat" and shallow minimum at $R=R_{eq}=12.54525$\,a.u. ,
see Table \ref{Ten2psu}, characterized by a small curvature. The depth of
the minimum is $0.0001215811$\,Ry, if it is compared to the asymptotics of the
potential curve (or, in other words, to the ground state energy of the Hydrogen
atom). The existence to this minimum, due to the Van der Waals attraction at large
distances between the Hydrogen atom and the proton,
was demonstrated time ago (see e.g. \cite{Peek:1969} and references therein).

\begin{center}
\begin{figure}
\begin{tabular}{cc}
\subfloat[]{\includegraphics[width=8cm]{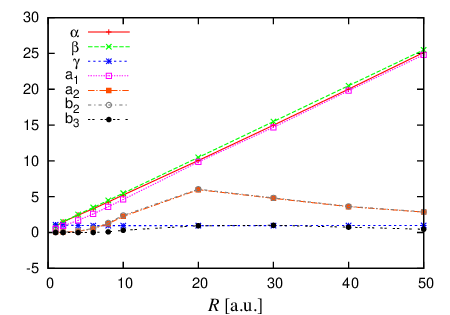}}
\subfloat[]{\includegraphics[width=8cm]{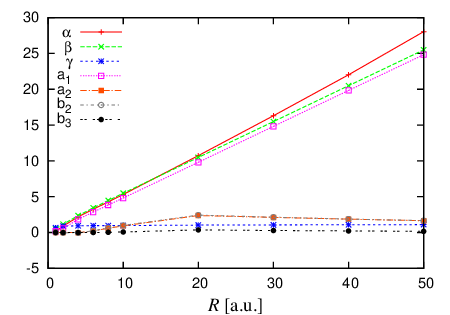}}
\end{tabular}
\caption{Variational parameters as a function of the internuclear
  distance $R$ for the two lowest states: (a)\ the positive parity
  state $1s\sigma_g$ and (b) the negative parity state $2p\sigma_u$.}
\label{figparms}
\end{figure}
\end{center}

Variational parameters are smooth, slow-changing functions of $R$,
see Figure \ref{figparms}. Note that the number of optimization parameters
can be reduced by putting $a_2=b_2=0$. In this case the accuracy in energy drops
from 10-11 to 5-6 s.d. - it is still acceptable from physics point of view being
consistent with a domain of applicability of non-relativistic QED, see a discussion
above. All calculations are implemented in double precision arithmetics and checked
in quadruple precision one.

Hence, our relatively-simple, few parametric functions (\ref{appr-0})
taken as trial functions in a variational study provide extremely high
accuracy in energy in comparison with highly-accurate alternative
calculations usually much more demanding computationally.
Two naturally related questions occur: (i) can we estimate the accuracy
of variationally obtained energies without making a comparison with other
calculations and (ii) how close locally our functions to the exact ones
in configuration space. In order to answer these questions, we develop
a perturbation theory in the Riccati equation derived from
the Schr\"odinger equation (\ref{Sch}) taking a trial function (\ref{appr-0})
as zero approximation, see \cite{Turbiner:2011}
\footnote{Such a procedure is called {\it no-linearization},
see \cite{Turbiner:1984}}.

\begin{center}
\begin{figure}
\begin{tabular}{cc}
\subfloat[]{\includegraphics[width=8cm]{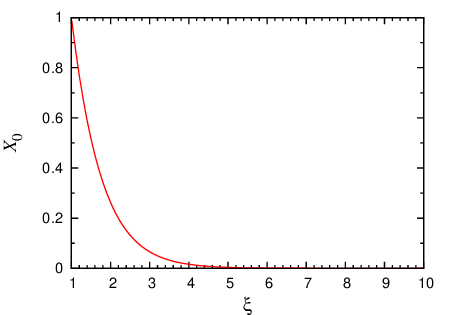}}
\subfloat[]{\includegraphics[width=8cm]{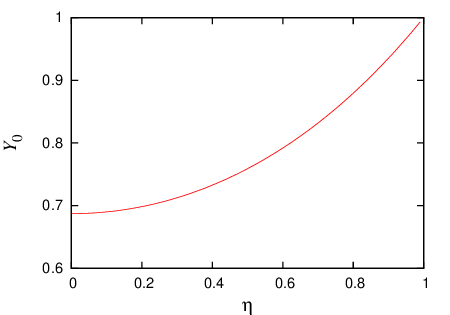}}
\end{tabular}
\caption{Components of the wave function for the ground state
  $1s\si_g$ for $R=2$~a.u.: (a) the $\xi$-dependent function $X_0$
  (\ref{appr-0}) and (b) the $\eta$-dependent function $Y_0^{(+)}$
  (\ref{appr-0}), cf. \cite{Turbiner:2011}.}
\label{fig-1ssg-XY}
\end{figure}
\end{center}

\begin{center}
\begin{figure}
\begin{tabular}{cc}
\subfloat[]{\includegraphics[width=8cm]{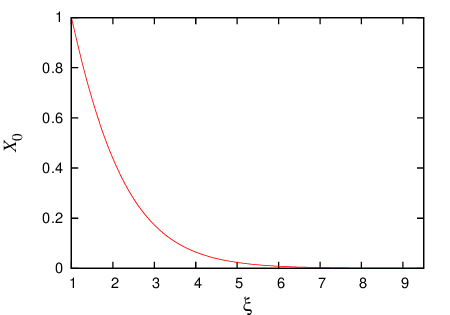}}
\subfloat[]{\includegraphics[width=8cm]{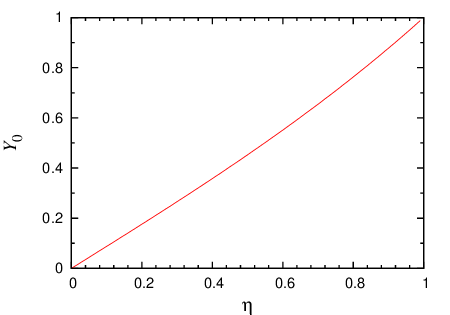}}
\end{tabular}
\caption{Components of the wave function for the first excited state
   $2p\si_u$ for $R=2$~a.u.: (a) the $\xi$-dependent function $X_0$
  (\ref{appr-0}) and (b) the $\eta$-dependent function $Y_0^{(-)}$
  (\ref{appr-0}), cf. \cite{Turbiner:2011}.}
\label{fig-2psu-XY}
\end{figure}
\end{center}

Let us choose $X_0, Y_0$ (\ref{appr-0}) with parameters fixed via
variational calculation (see above) as zero approximation in perturbation
theory (\ref{PTx}), (\ref{PTy}) (see Appendix). For given $X_0, Y_0$ one can
calculate a potential $V_0$, for which the approximation (\ref{appr-0})
is the exact eigenfunction. It is evident that by construction of $X_0, Y_0$
the emerging perturbation theory has to be convergent: the perturbation
potential $(V-V_0)$ is subdominant. Assuming the consistency condition (\ref{An})
is fulfilled for the first corrections, namely, $A_{1,\xi}\ =\ A_{1,\eta}=A_1$,
we find the first corrections $\varphi_1(\xi)$ and $\varrho_1(\eta)$ as functions
of $A_1$. Then we modify the trial function (\ref{appr-0}) accordingly,
\begin{equation}
\label{appr-1}
\Psi^{(\pm)}_{0,0,0} \rar X_0(\xi) Y_0^{(\pm)}(\eta)
\ e^{-\varphi_1(\xi)-\varrho_1(\eta)}
\end{equation}
and make the variational calculation with this trial function
minimizing with respect to parameter $p$. The (expected) result is
that the optimal value of parameter $p$ remained unchanged with
respect to the value obtained for the trial function (\ref{appr-0})
with extremely high accuracy - within 10 s.d.! It indicates that the
condition (\ref{An}) is fulfilled with high accuracy. The variational
energy is changed beyond the 10 s.d. Therefore, our energies presented
in Tables \ref{Ten1ssg}, \ref{Ten2psu} are correct in all ten digits. The
separation parameters $A_{1,\xi}, A_{1,\eta}$ are presented in Table
\ref{Aval} together with those corresponding to other states (see
below). It allows us to find explicitly $\varphi_1(\xi)$ and
$\varrho_1(\eta)$. As an illustration the functions
$X_0(\xi)$ and $Y_0^{(\pm)}(\eta)$ are shown for $R=2$ a.u. on
Figs. \ref{fig-1ssg-XY} and \ref{fig-2psu-XY}.
Corrections $\varphi_1(\xi)$ and $\varrho_1(\eta)$ are shown in~\cite{Turbiner:2011}.
Similar behavior of corrections appears for all other values of $R \in [0, 40]$.

\begin{table}[ht]
\begin{center}
\caption{The total energy $E(R)$ for $1s \si_g$ state of the H$_2^+$-ion compared to \cite{Montgomery:1977} (rounded) and Lagrange mesh method (cf. \cite{Turbiner:2011}); $R_{eq}=1.997193$\,a.u.\ , $p$ is taken as variational parameter}
\label{Ten1ssg}
\begin{tabular}{ccc}
\hline \hline
R[a.u.] & $E$[Ry] (Present/\cite{Montgomery:1977}/Mesh)
        &\qquad $p$ \hspace{0.9cm} \phantom{.}\\
\hline \hline
1.0     &-0.90357262676  &\,\,0.8519936\\
        &-0.90357262676  & \\
        &-0.90357262676  & \\[5pt]
\br
1.997193& -1.20526923821   &\,\,1.483403 \\
        & \qquad --        & \\
        & -1.20526923821   &\\[5pt]
\br
2.0     &-1.20526842899  &\,\,1.485015\\
        &-1.20526842899  & \\
        &-1.20526842899  & \\[5pt]
\br
40.0   &-1.0000017622   & 20.4939\\
       &\,\,\,-----     &\\
       &-1.0000017622   &\\
\hline \hline
\end{tabular}
\end{center}
\end{table}

\begin{table}[ht]
\begin{center}
\caption{The total energy $E(R)$ for $2p \si_u$ state of the H$_2^+$-ion compared to \cite{Montgomery:1977} (rounded) and Lagrange mesh method (cf. \cite{Turbiner:2011});
$R_{eq}=12.54525$\,a.u.\ ,
$p$ is taken as variational parameter}
\label{Ten2psu}
\begin{tabular}{ccc}
\hline \hline
R\ [a.u.] &\phantom{.}\, $E$\ (Present/\cite{Montgomery:1977}/Mesh)\ [Ry]
          &\phantom{.} \qquad $p$ \hspace{0.9cm} \phantom{.}\\
\hline \hline
1.0     &\,\,0.8703727499   & 0.5314196\\
        &\,\,0.8703727498   & \\
        &\,\,0.8703727498   & \\[5pt]
\br
1.997193 & -0.3332800331   & 1.1536645\\
         & \,\,\,-----     & \\
         & -0.33328003316  & \\[5pt]
12.54525 & -1.0001215811   & 6.75434 \\
         & \,\,\,---       & \\
         & -1.0001215811   & \\[5pt]
\br
40.0     &-1.0000017622   & 20.4939 \\
         &\,\,\,---        &\\
         &-1.0000017622   &\\
\hline \hline
\end{tabular}
\end{center}
\end{table}

\subsection{$(0,0,\La,\pm)$ states with $\La=1,2$}

As the further check the quality of the approximation \re{appr}
proposed in \cite{Turbiner:2011}, we considered the states with
magnetic quantum number $\La=1,2$ and both parities $(\pm)$. These
four states $(0,0,1,+)$, $(0,0,1,-)$, $(0,0,2,+)$ and $(0,0,2,-)$
correspond to the states $2p\pi_u$, $3d\pi_g$, $3d\de_g$ and $4f\de_u$
in the united atom nomenclature, respectively. The approximation takes
the form
\begin{equation}
\label{apprL}
 \Psi^{(\pm)}_{0,0,\La}=\frac{(\xi^2-1)^{\La/2}e^{-\xi\frac{\al+p\xi}{\gamma+\xi}}}
 {(\gamma+\xi)^{1+\La-\frac{R}{p}}}
 \frac{(1-\eta^2)^{\La/2}}{(1 + b_2 \eta^2 + b_3 \eta^4)^{\frac{1+\La}{4}}}
 \left[
   \begin{array}{c}
   \cosh\\
   \sinh
   \end{array}
 \left(\eta\frac{a_1+p a_2\eta^2 +p b_3\eta^4}{1+b_2\eta^2+b_3\eta^4}\right)
 \right]
  e^{\pm i \La \phi} \ ,
\end{equation}
for positive and negative parity, respectively; it depends on six free
parameters $\al,\ga$ and $a_{1,2}$, $b_{2,3}$ as well as $p$ which can
be taken as an extra variational parameter. Due to the presence of the
last factor in $\Psi^{(\pm)}_{0,0,\La}$ the function (\ref{apprL}) is
orthogonal to (\ref{appr-0}). Taking \re{apprL} as a trial function
and using the variational method, the optimized values of these
parameters are obtained for each fixed value of the internuclear
distance $R$.  The results for the total energy and the value of p for
the states with $\La=1, 2$ and both parities $(\pm)$ as a function of
the internuclear distance $R$ are presented in Table
\ref{tb00Lpm}. For each $R$-value, the second line are the results
presented by Madsen and Peek~\cite{Mar:1971}. In general, the
agreement is on the level of $10$ s.d. except for a few values of $R$
where the agreement is on $8-9$ s.d.
It can be clearly seen the pairing phenomenon, see Table
\ref{tb00Lpm}: energies of the states of different parities approach
to each other with growth of $R$. At $R=50$\,a.u. the energy gaps reach
$\sim 10^{-8}$\,Ry and $\sim 10^{-5}$\,Ry for $(0,0,1,\pm)$ and $(0,0,2,\pm)$
states, respectively. The energy difference with appropriate state of
Hydrogen atom,
which occurs after dissociation, at $R\rar \infty$, see e.g. Table~\ref{Rto0},
is $\sim 10^{-3}$\,Ry for $R=50$\,a.u.
This difference reduces gradually with further growth of $R$.

\begin{center}
\begin{figure}
\begin{tabular}{cc}
\subfloat[]{\includegraphics[width=8cm]{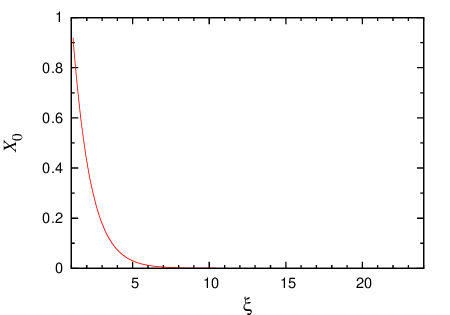}}
\subfloat[]{\includegraphics[width=8cm]{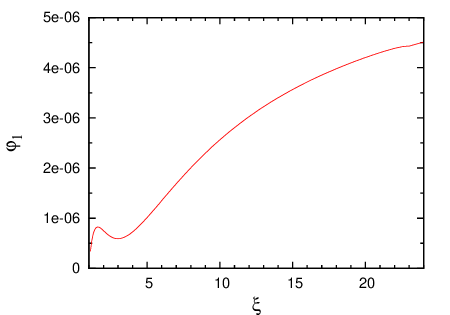}}
\end{tabular}
\caption{The $2p\pi_u$ state at $R=2$ a.u.: (a)\ $\xi$-dependent function $X_0$ (\ref{apprL}) and (b)\ the
first correction $\varphi_1$ (see e.g. (\ref{apprLC})).}
\label{fig-2ppu-X}
\end{figure}
\end{center}

\begin{center}
\begin{figure}
\begin{tabular}{cc}
\subfloat[]{\includegraphics[width=8cm]{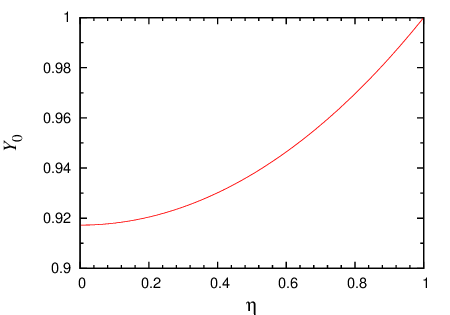}}
\subfloat[]{\includegraphics[width=8cm]{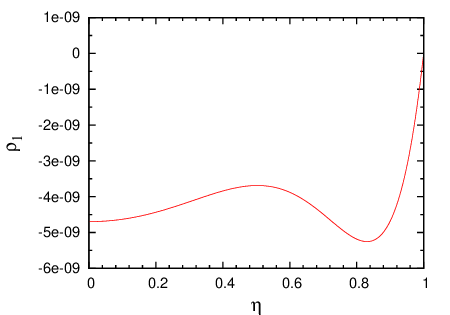}}
\end{tabular}
\caption{The $2p\pi_u$ state at $R=2$ a.u.: (a)\ $\eta$-dependent function $Y_0^{(-)}$ \re{apprL} and (b)\
the first correction $\rho_1$ (see e.g. \re{apprLC}).}
\label{fig-2ppu-Y}
\end{figure}
\end{center}

\begin{center}
\begin{figure}
\begin{tabular}{cc}
\subfloat[]{\includegraphics[width=8cm]{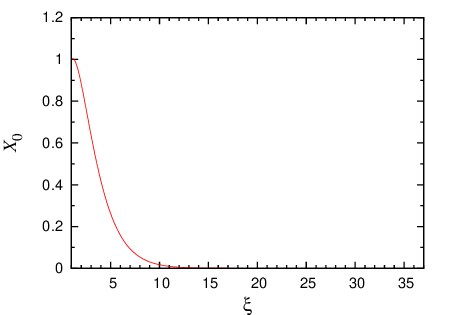}}
\subfloat[]{\includegraphics[width=8cm]{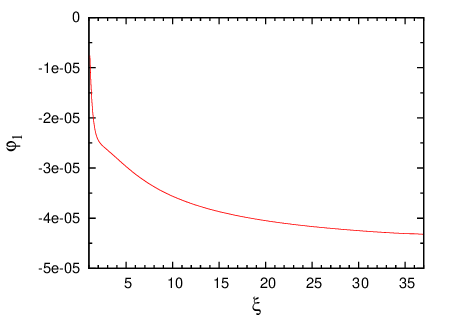}}
\end{tabular}
\caption{The $3d\pi_g$ state at $R=2$ a.u.: (a)\ $\xi$-dependent function $X_0$ \re{apprL} and (b)\ the
first correction $\varphi_1$ (see e.g. \re{apprLC}).}
\label{fig-3dpg-X}
\end{figure}
\end{center}

\begin{center}
\begin{figure}
\begin{tabular}{cc}
\subfloat[]{\includegraphics[width=8cm]{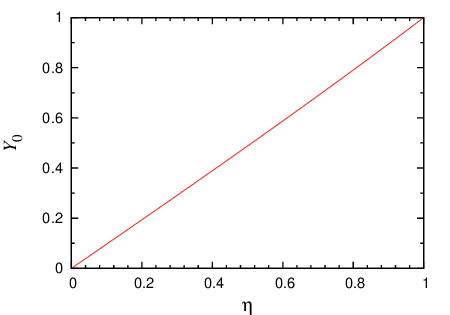}}
\subfloat[]{\includegraphics[width=8cm]{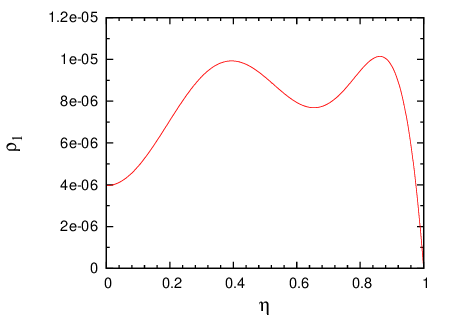}}
\end{tabular}
\caption{The $3d\pi_g$ state at $R=2$ a.u.: (a)\ $\eta$-dependent function $Y_0^{(-)}$ \re{apprL} and (b)\
the first correction $\rho_1$ (see e.g. \re{apprLC}).}
\label{fig-3dpg-Y}
\end{figure}
\end{center}

\begin{center}
\begin{figure}
\begin{tabular}{cc}
\subfloat[]{\includegraphics[width=8cm]{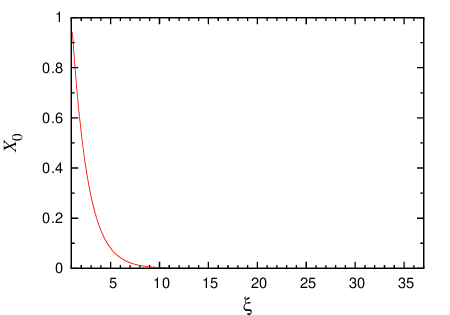}}
\subfloat[]{\includegraphics[width=8cm]{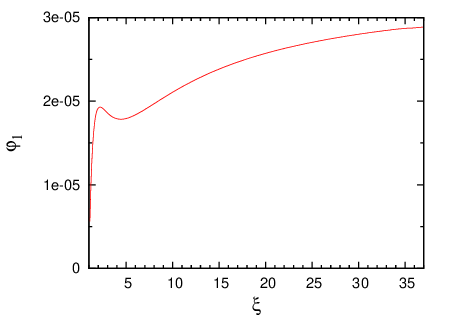}}
\end{tabular}
\caption{The $3d\de_g$ state at $R=2$ a.u.: (a)\ $\xi$-dependent function $X_0$ \re{apprL} and (b)\ the
first correction $\varphi_1$ (see e.g. \re{apprLC}).}
\label{fig-3ddg-X}
\end{figure}
\end{center}

\begin{center}
\begin{figure}
\begin{tabular}{cc}
\subfloat[]{\includegraphics[width=8cm]{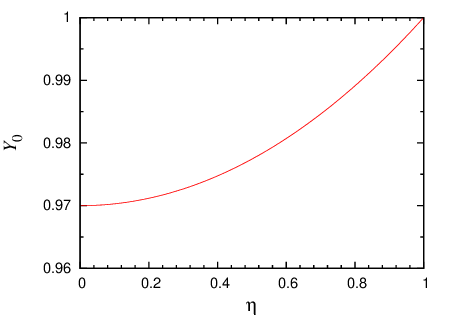}}
\subfloat[]{\includegraphics[width=8cm]{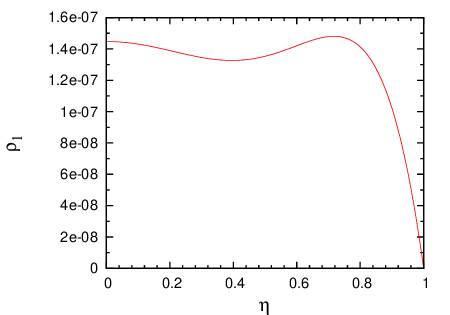}}
\end{tabular}
\caption{The $3d\de_g$ state at $R=2$ a.u.: (a)\ $\eta$-dependent function $Y_0^{(-)}$ \re{apprL} and
(b)\ the first correction $\rho_1$ (see e.g. \re{apprLC}).}
\label{fig-3ddg-Y}
\end{figure}
\end{center}

\begin{center}
\begin{figure}
\begin{tabular}{cc}
\subfloat[]{\includegraphics[width=8cm]{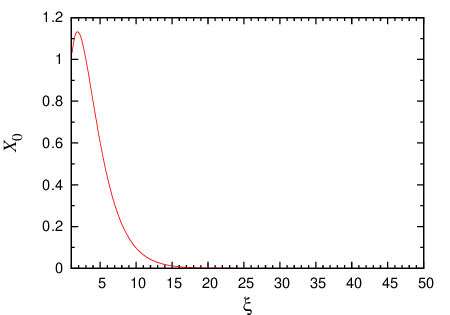}}
\subfloat[]{\includegraphics[width=8cm]{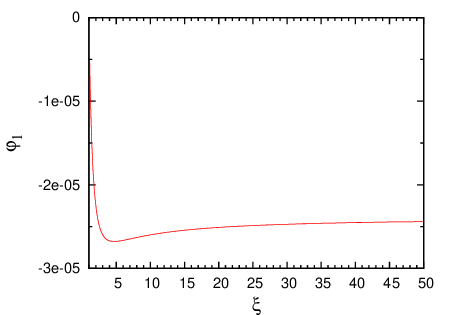}}
\end{tabular}
\caption{The $4f\de_u$ state at $R=2$ a.u.: (a)\ $\xi$-dependent function $X_0$ \re{apprL} and (b)\ the
first correction $\varphi_1$ (see e.g. \re{apprLC}).}
\label{fig-4fdu-X}
\end{figure}
\end{center}

\begin{center}
\begin{figure}
\begin{tabular}{cc}
\subfloat[]{\includegraphics[width=8cm]{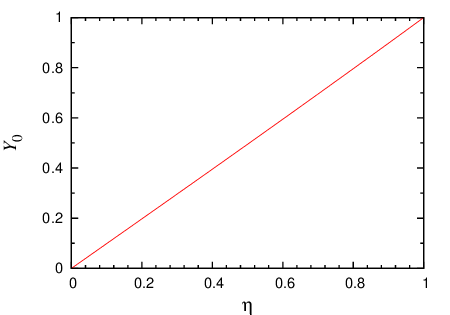}}
\subfloat[]{\includegraphics[width=8cm]{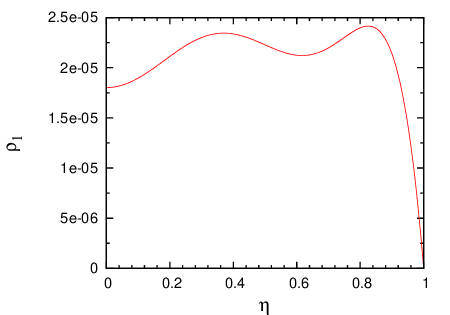}}
\end{tabular}
\caption{The $4f\de_u$ state at $R=2$ a.u.: (a)\ $\eta$-dependent function $Y_0^{(-)}$ \re{apprL} and
(b)\ the first correction $\rho_1$ (see e.g. \re{apprLC}).}
\label{fig-4fdu-Y}
\end{figure}
\end{center}

In a similar way how it was done for ground states of the positive and negative parities
(\ref{appr-0}) one can develop convergent perturbation theory for $(0,0,\La,\pm)$ states
individually, see Appendix, taking (\ref{apprL}) as a zero approximation. Taking into
account the first-order corrections $\varphi_1, \varrho_1$ to the function (\ref{apprL}),
it gets the form
\begin{equation}
\label{apprLC}
\Psi^{(\pm)}_{0,0,\La} \rar X_0(\xi) Y_0^{(\pm)}(\eta)\,  e^{\pm i \La \phi} \,
     e^{-\varphi_1(\xi)-\varrho_1(\eta)}\ .
\end{equation}
The immediate striking observation is the first correction to energy is of order
$10^{-11}$ (or less) and influences digits beyond those shown in Table~\ref{tb00Lpm}!
On Figs. \ref{fig-2ppu-X}, \ref{fig-2ppu-Y}, \ref{fig-3dpg-X},
\ref{fig-3dpg-Y}, \ref{fig-3ddg-X}, \ref{fig-3ddg-Y}, \ref{fig-4fdu-X}
and \ref{fig-4fdu-Y} the trial functions $X_0(\xi)$,
$Y_0^{(\pm)}(\eta)$ and the first corrections to the phases
$\varphi_1(\xi)$ and $\varrho_1(\eta)$ for $R=2$ a.u. are present. We
must emphasize that the variational parameter $p$ in Table
\ref{tb00Lpm} coincides with the value of $p$  \textcolor{blue}{(\ref{p})}
found from the variational energy, on the level of 5 - 9 s.d. It indicates
implicitly to a very high quality of the function (\ref{apprL}).

\begin{sidewaystable}
\caption{Total energy $E(R)$ for the $(0,0,\La=1,2,\pm)$ states of the H$_2^+$-ion
compared to \cite{Mar:1971} (second line).}
\label{tb00Lpm}
{\scriptsize
\begin{center}
\begin{tabular}{l|ll|ll|ll|ll}
\hline \hline
 R[a.u.] &$E(0,0,1,+)$&$p(0,0,1,+)$&$E(0,0,1,-)$&$p(0,0,1,-)$&$E(0,0,2,+)$&$p(0,0,2,+)$&$E(0,0,2,-)$&
 $p(0,0,2,-)$\\
\hline \hline
1.0 &\m1.051\,784\,087\,48(77)&0.486\,882 &\m1.552\,886\,885\,83 &0.334\,332\,5&  1.560\,917\,409\,665  &0.331\,316\,5&\m1.750\,004\,925\,60 &0.249\,9975\\
    &\m1.051\,784\,087\,4746& &\m1.552\,886\,885\,8238& & 1.560\,917\,409\,6654& &\m1.750\,004\,925\,5960& \\
2.0 &\m0.142\,456\,360\,21(08)&0.926\,037 &\m0.546\,600\,746\,72 &0.673\,349   &  0.574\,534\,636\,379  &0.652\,277&\m0.750\,074\,914\,13   &0.4999\,25\\
    &\m0.142\,456\,360\,20826 &          &\m0.546\,600\,746\,7126&             & 0.574\,534\,636\,3784  &           &\m0.750\,074\,914\,1264 &           \\
4.0 &-0.201\,649\,288\,23     &1.675\,29  &\m0.038\,093\,115\,38 &1.359\,274\,6& 0.111\,109\,971\,587   &1.247\,221&\m0.250\,988\,746\,1   &0.998\,02\\
    &-0.201\,649\,288\,2302   &          &\m0.038\,093\,115\,3803&             & 0.111\,109\,971\,58626 &           &\m0.250\,988\,746\,0990 &           \\
6.0 &-0.260\,649\,791\,31     &2.312\,11 &-0.121\,744\,444\,93   &2.0237\,84   &-0.019\,437\,228\,851   &1.781\,835&\m0.087\,106\,783\,3    &1.488\,64\\
    &-0.260\,649\,791\,3114   &          &-0.121\,744\,444\,95100&             &-0.019\,437\,228\,851128&           &\m0.087\,106\,783\,24228&           \\
8.0 &-0.269\,021\,262\,54(37) &2.881\,725&-0.188\,783\,036\,57   &2.649\,628   &-0.071\,453\,569\,562   &2.267\,87502&\m0.008\,575\,937\,0    &1.965\,4\\
    &-0.269\,021\,262\,5382   &          &-0.188\,783\,036\,58772&             &-0.071\,453\,569\,562040&           &\m0.008\,575\,936\,87662&           \\
10.0&-0.265\,432\,580\,28     &3.411\,13 &-0.219\,833\,749\,01   &3.239\,73    &-0.095\,093\,601\,175   &2.716\,126& -0.035\,171\,033\,8    &2.424\,722\\
    &-0.265\,432\,580\,2914   &          &-0.219\,833\,749\,0582 &             &-0.095\,093\,601\,17488 &           & -0.035\,171\,034\,00198&           \\
14.0&-0.255\,396\,545\,98     &4.417\,514&-0.242\,319\,086\,11   &4.344\,4     &-0.111\,495\,667\,118   &3.530\,34& -0.078\,059\,693\,0    &3.2901\,3\\
    &-0.255\,396\,546\,2922   &          &-0.242\,319\,086\,1426 &             &-0.111\,495\,667\,11852 &           & -0.078\,059\,693\,28034&           \\
20.0&-0.250\,167\,097\,81     &5.917\,5  &-0.248\,752\,926\,71   &5.905\,531   &-0.113\,758\,110\,521   &4.623\,4& -0.100\,623\,878\,9    &4.4791\,1\\
    &-0.250\,167\,098\,9774   &          &-0.248\,752\,926\,741  &             &-0.113\,758\,110\,6214  &           & -0.100\,623\,879\,14096&           \\
30.0&-0.249\,755\,905\,14     &8.437\,72 &-0.249\,734\,619\,46   &8.437\,4     &-0.110\,960\,223\,7   &6.322& -0.109\,116\,195\,2    &6.28897\\
    &-0.249\,755\,905\,4846   &          &-0.249\,734\,619\,4714 &             &-0.110\,960\,225\,79684 &           & -0.109\,116\,195\,34154&           \\
40.0&-0.249\,872\,858\,88     &10.952\,1 &-0.249\,872\,610\,06   &10.952\,1    &-0.110\,588\,155   &8.014\,7& -0.110\,429\,620\,8    &8.01073\\
    &-0.249\,872\,859\,9708   &          &-0.249\,872\,610\,0936 &             &-0.110\,588\,156\,5852  &           & -0.110\,429\,620\,89928&           \\
50.0&-0.249\,928\,750\,05     &13.461\,3 &-0.249\,928\,747\,50   &13.461\,26   &-0.110\,756\,575   &9.706\,8& -0.110\,745\,830\,0    &9.7065\\
    &-0.249\,928\,750\,0956   &          &-0.249\,928\,747\,5080 &             &-0.110\,756\,576\,55914 &           & -0.110\,745\,830\,06112&           \\
\hline \hline
\end{tabular}
\end{center}}
\end{sidewaystable}

\subsection{Ellipsoidal nodal surfaces: the $(1,0,0,\pm)$ states}

The proposed approximation~\re{appr} \cite{Turbiner:2011} allows us to
study the $n$th excited state in $\xi$ direction with $n$ nodes in the
$\xi$ variable. Let us consider the simplest states, $n=1$, $\La=0$
and both parities $(\pm)$, {\it i.e.} the states $(1,0,0,\pm)$ or,
differently, $2s\si_g$ and $3p\si_u$, respectively. The main
difference with the approximation for the ground state \re{appr-0}
comes due to the presence of a monomial factor $(\xi-\xi_0)$ in the
expression for $X_0(\xi)$, while the $Y_0(\eta)$ remains functionally
the same,
\begin{equation}
\label{apprNX}
 X_{0} = \frac{(\xi-\xi_0)}{(\gamma + \xi)^{2-\frac{R}{p}}}
 e^{-\xi \frac{\al + p \xi}{\gamma + \xi}}\ .
\end{equation}
Here $\xi_0$ defines the position of the node and it can be fixed by
imposing the orthogonality condition between these states ($\pm$
parity) and the lowest states, {\it i.e.}
$\langle(0,0,0,\pm)|(1,0,0,\pm)\rangle = 0$.  The orthogonality with
the states $(0,0,\La,\pm)$ for any $\La$ is always
fulfilled. Eventually, the approximation $\Psi^{(\pm)}_{1,0,0}$
contains six free parameters which are obtained using the variational
method. Results are presented in Table~\ref{tb10Lpm} for the two
states $2s\si_g$ $(1,0,0,+)$ and $3p\si_u$ $(1,0,0,-)$ as a function
of the internuclear distance $R$.  Comparison with previous, highly
accurate results~\cite{Mar:1971} (given on the second line) for each
$R$ is presented.  The agreement is on the level of $10$
s.d. For each state the variational value of $p$ (when $p$ is taken as
a variational parameter in (\ref{appr})) as well as the node position
$\xi^{\pm}_0$ are given.
It can be clearly seen on Table \ref{tb10Lpm} the pairing phenomenon:
energies of the states of opposite parities approach to each other
with growth of $R$. At $R=40$\,a.u. the energy gap reaches $\sim
10^{-8}$\,a.u. for $(1,0,0,\pm)$ states. The energy difference with
appropriate state of Hydrogen atom is $\sim 10^{-2}$\,a.u., see
e.g. Table~\ref{Rto0}, for the case $R \rar \infty$
this difference should reduce gradually with further growth of $R$.

\begin{center}
\begin{figure}
\begin{tabular}{cc}
\subfloat[]{\includegraphics[width=8cm]{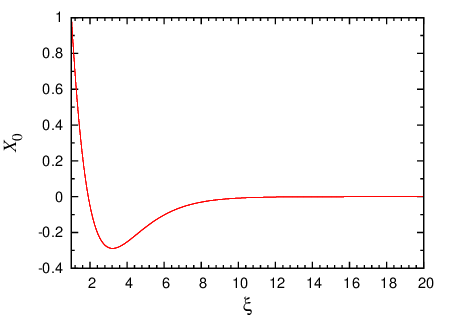}}
\subfloat[]{\includegraphics[width=8cm]{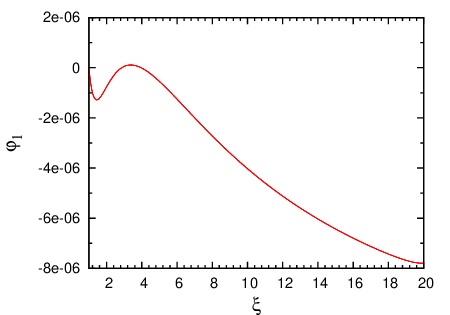}}
\end{tabular}
\caption{The $2s\si_g$ state at $R=2$ a.u.: (a)\ $\xi$-dependent function $X_0$ \re{apprNX} and (b)\ the
first correction $\varphi_1$ (see e.g. \re{appr-1}).}
\label{fig-2ssg-X}
\end{figure}
\end{center}

\begin{center}
\begin{figure}
\begin{tabular}{cc}
\subfloat[]{\includegraphics[width=8cm]{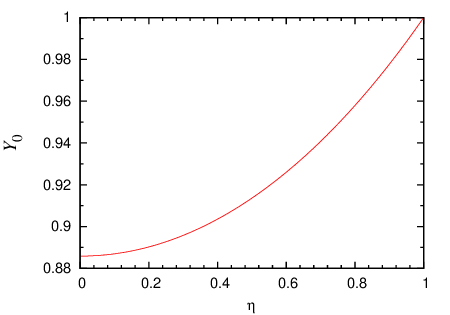}}
\subfloat[]{\includegraphics[width=8cm]{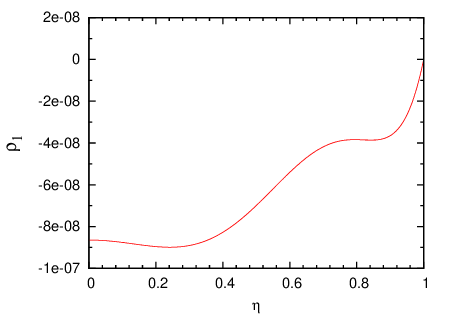}}
\end{tabular}
\caption{The $2s\si_g$ state at $R=2$ a.u.: (a)\ $\eta$-dependent function $Y_0^{(-)}$ \re{appr-0} and
(b)\ the first correction $\rho_1$ (see e.g. \re{appr-1}).}
\label{fig-2ssg-Y}
\end{figure}
\end{center}

\begin{center}
\begin{figure}
\begin{tabular}{cc}
\subfloat[]{\includegraphics[width=8cm]{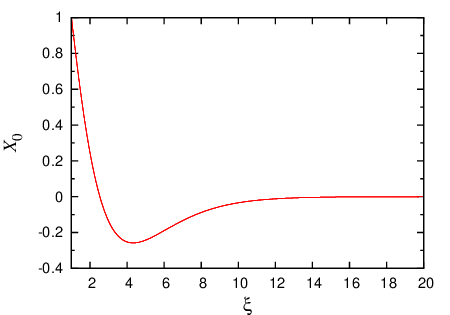}}
\subfloat[]{\includegraphics[width=8cm]{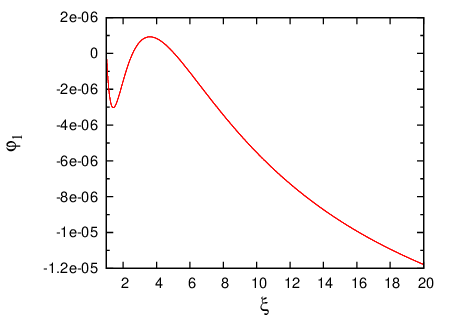}}
\end{tabular}
\caption{The $3p\si_u$ state at $R=2$ a.u.: (a)\ $\xi$-dependent function $X_0$ \re{apprNX} and (b)\ the
first correction $\varphi_1$ (see e.g. \re{appr-1}).}
\label{fig-3psu-X}
\end{figure}
\end{center}

\begin{center}
\begin{figure}
\begin{tabular}{cc}
\subfloat[]{\includegraphics[width=8cm]{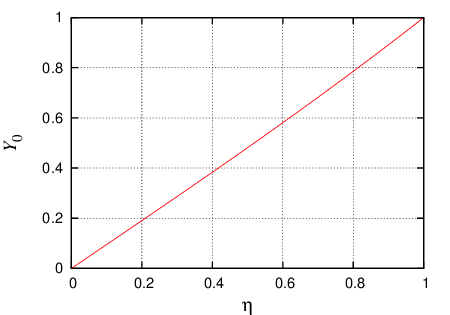}}
\subfloat[]{\includegraphics[width=8cm]{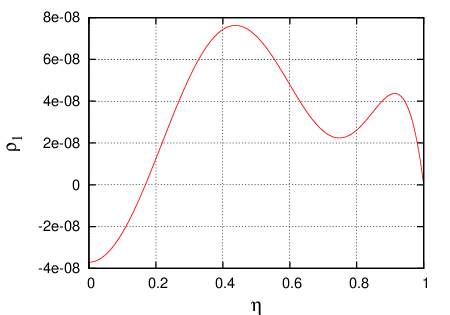}}
\end{tabular}
\caption{The $3p\si_u$ state at $R=2$ a.u.: (a)\ $\eta$-dependent function $Y_0^{(-)}$ \re{appr-0} and
(b)\ the first correction $\rho_1$ (see e.g. \re{appr-1}).}
\label{fig-3psu-Y}
\end{figure}
\end{center}

In both cases $(1,0,0,\pm)$ the node position is a decreasing function
of the internuclear distance having a finite value for small $R$ and
conversely approaching to the lower limit in $\xi$-coordinate, $\xi =
1$ at large $R$, roughly as $\sim 1/R$.  At the point $\xi^{\pm}_0$,
the wave function (\ref{apprNX}) vanishes. In the configuration space
it corresponds to a nodal surface which is a prolate spheroid of
eccentricity $\varepsilon = 1/\xi^{\pm}_0$. Corrections to the
node-position can be calculated developing a convergent perturbation
theory (see Appendix, Eq.\re{xf_n}). We found that for these two
states the first correction is $\sim 10^{-7}$.  Functions $X_0(\xi)$,
$Y_0^{(\pm)}(\eta)$ and the first corrections to the phases are shown
in Figs. \ref{fig-2ssg-X}, \ref{fig-2ssg-Y}, \ref{fig-3psu-X} and
\ref{fig-3psu-Y} for $R=2$ a.u. as an illustration.

\begin{table}[!th]
\begin{center}
  \caption{Total energy $E(R)$ for the $2s\si_g$ $(1,0,0,+)$ and
    $3p\si_u$ $(1,0,0,-)$ states (left/right columns, respectively)
    of the H$_2^+$ molecular ion (the first line) compared to \cite{Mar:1971}
    (the second line). $\xi^{\pm}_0$ gives the node position.}
\label{tb10Lpm}
\begin{tabular}{l|lll|lll}
\hline \hline
 R[a.u.] &\ $E(1,0,0,+)$\ &\ $p(1,0,0,+)$&$\ \xi^{+}_0$&\ $E(1,0,0,-)$&\ $p(1,0,0,-)$&$\ \xi^{-}_0$\\
\hline \hline
1.0 &\m1.154\,150\,822\,6     &  0.4598503&2.782853311&\m1.521\,369\,039\,285  & 0.345916  &5.360475264\\
    &\m1.154\,150\,823\,003   &	          &	        &\m1.521\,369\,039\,2720 &           &           \\
2.0 &\m0.278\,270\,249\,325   &  0.849547 &1.907869613&\m0.489\,173\,669\,829  & 0.714721  &2.532742379\\
    &\m0.278\,270\,249\,323\,4&           &           &\m0.489\,173\,669\,8286 &           &           \\
4.0 &-0.077\,029\,734\,913\,5 &  1.5192495&1.477672193&\m0.009\,780\,899\,90   & 1.40031296&1.589362953\\
    &-0.077\,029\,734\,914\,98&           &           &\m0.009\,780\,899\,904368&           &           \\
6.0 &-0.161\,775\,845\,624    &  2.11092  &1.330973187&-0.121\,531\,762\,3     & 2.02331   &1.364704127\\
    &-0.161\,775\,845\,629\,74&           &           &-0.121\,531\,762\,33782 &           &           \\
8.0 &-0.193\,554\,665\,734    &  2.663996 &1.254298836&-0.174\,967\,289\,16    & 2.60758   &1.265974957\\
    &-0.193\,554\,665\,735\,18&	          &           &-0.174\,967\,289\,19184 &           &           \\
10.0&-0.209\,421\,251\,79     &  3.1993   &1.206019531&-0.201\,171\,505\,95    & 3.16691   &1.210160770\\
    &-0.209\,421\,251\,818\,4 &           &           &-0.201\,171\,506\,037   &           &           \\
20.0&-0.236\,998\,606\,92     &  5.80516  &1.103266490&-0.236\,904\,750\,195   & 5.80435   &1.103289607\\
    &-0.236\,998\,606\,945\,2 &           &           &-0.236\,904\,750\,2114  &           &           \\
30.0&-0.243\,892\,622\,63     &  8.35918  &1.068352565&-0.243\,891\,770\,96    & 8.35916   &1.068352748\\
    &-0.243\,892\,622\,973\,6 &           &           &-0.243\,891\,770\,9742  &           &           \\
40.0&-0.246\,478\,659\,89     & 10.88997  &1.051017992&-0.246\,478\,652\,70    &10.88997   &1.051017930\\
    &-0.246\,478\,659\,911\,8 &           &           &-0.246\,478\,652\,7404  &           &           \\
50.0&-0.247\,714\,222\,867    & 13.40975  &1.040679396&-0.247\,714\,222\,80    &13.40975   &1.040679432\\
    &-0.247\,714\,222\,873\,8 &           &           &-0.247\,714\,222\,8160  &           &           \\
\hline \hline
\end{tabular}
\end{center}
\end{table}

\subsection{$(0,1,0,\pm)$ states }

Now, let us consider states with two nodes in the $\eta$-coordinate at
$\Lambda=0$. These states correspond to $(0,1,0,\pm)$ or in the
united atom notation $3d\sigma_g$ and $4f\sigma_u$, respectively. The
functional form of the $X(\xi)$ function is the same as one of the
ground state~\re{appr-0} while the $Y(\eta)$ function is given by
\begin{equation}
\label{eta-state}
 Y(\eta) = \frac{(\eta^2-\eta_0^2)}{(1 + b_2 \eta^2 + b_3 \eta^4)^{1/4}}
 \left[
   \begin{array}{c}
    \cosh  \\
    \sinh
    \end{array}
 \left(\eta \frac{a_1 + p a_2 \eta^2 + p b_3 \eta^4}
 {1 + b_2\eta^2 + b_3 \eta^4}\right) \right].
\end{equation}
This contain a second-degree polynomial $(\eta^2-\eta_0^2)$
(cf.~\re{appr}) indicating the node positions $\pm\eta_0$ which are fixed
by the orthogonality condition to the states $(0,0,0,\pm)$. The
approximation $\Psi^{(\pm)}_{0,1,0}$ contains six free parameters whose
are going to be optimized by applying the variational method.

Table~\ref{tb01Lpm} presents the results for the total energy $E$ as
well as the values of the $p$-parameter and the node position
$\eta=\eta_{0^{\pm}}$. The nodes appear symmetrically with
respect to $\eta=0$. The node surfaces are hyperboloids of revolution
around the internuclear axis with eccentricity $\varepsilon=
1/\eta_{0^{\pm}}$.  One can see in Table~\ref{tb01Lpm} there is a
dramatic decrease in the accuracy of the variational energies of both
states for small internuclear distances. When comparing the total
energy of the state $(0,1,0,+)$ with the results presented by Madsen
and Peek \cite{Mar:1971} (second row) we have 7 s.d. in agreement for
$R\in[6 - 50]$~a.u. dropping steadily to 4 s.d. for $R \in [1 -
4]$~a.u. With regard to the state $(0,1,0,-)$ the agreement is in 8-9
s.d. for $R\in[20 - 50]$~a.u. decreasing to 3-4 s.d. for $R\in[1 -
10]$~a.u.
It can be clearly seen on Table \ref{tb01Lpm} the pairing phenomenon:
energies of the states of different parities approach to each other
with growth of $R$. At $R=50$\,a.u. the energy gap reaches $\sim
10^{-7}$\,Ry for $(0,1,0,\pm)$ states. The energy difference with
appropriate state of the Hydrogen atom at $R\rar \infty$ is $\sim
10^{-3}$\,Ry, this difference should reduce gradually with further
growth of $R$.

Calculation of the 2nd correction to energy (and the first
correction to the node positions) with Eq.~\re{yg_n}, see Appendix,
does not improve significantly the variational energies. The first
correction to $\eta_{0}$ is very small.  This is the indication to a
slower convergence of the perturbation theory for those states
compared to other states. It is evident that the pre-factor in
(\ref{eta-state}), which describes nodes in $\eta$ (simple zeroes),
should be more complicated than simply $(\eta^2-\eta_0^2)$.
For the moment, it is not clear in what direction it has to be modified.

\begin{table}[!th]
\begin{center}
  \caption{Total energy $E(R)$ in Ry for the $3d\si_g$ $(0,1,0,+)$ and
    $4f\si_u$ $(0,1,0,-)$ states of the H$_2^+$ molecular ion (the
    first line) compared to \cite{Mar:1971} (the second
    line). $\eta_{0^{\pm}}$ in [a.u.] gives the node position.}
\label{tb01Lpm}
\begin{tabular}{l|lll|lll}
\hline \hline
 R[a.u.] &$E(0,1,0,+)$&$p(0,1,0,+)$&$\eta^{2}_{0^+}$&$E(0,1,0,-)$&$p(0,1,0,-)$&$\eta^{2}_{0^-}$\\
\hline \hline
1.0 &\m1.549\,645        & 0.33555 &0.33559 & 1.749\,201            & 0.2504&0.7750\\
    &\m1.549\,630\,623873&         &        & 1.749\,199\,647\,6496 &       &      \\
2.0 &\m0.528\,467        & 0.6867  &0.34330 & 0.746\,725            & 0.50328&0.7761\\
    &\m0.528\,444\,742349&         &        & 0.746\,712\,259\,7008 &        &      \\
4.0 &-0.071\,447\,03     & 1.51188 &0.385086& 0.235\,14             & 1.0294&0.7813\\
    &-0.071\,447\,5809595&         &        & 0.235\,095\,441\,5056 &       &      \\
6.0 &-0.291\,656\,30     & 2.37169 &0.467547& 0.041\,358            & 1.621&0.7918\\
    &-0.291\,656\,3202834&         &        & 0.041\,339\,661\,36794&       &      \\
8.0 &-0.347\,023\,26    & 3.0907   &0.55840 &-0.085\,484\,8         & 2.31685&0.8084\\
    &-0.347\,023\,2833194&         &        &-0.085\,486\,025\,27608&        &       \\
10.0&-0.346\,234\,878    & 3.6954  &0.633665&-0.169\,773\,24        & 3.04045&0.828782\\
    &-0.346\,234\,8809774&         &        &-0.169\,773\,328\,91252&        &        \\
20.0&-0.275\,532\,157    & 6.128   &0.813436&-0.261\,170\,845\,7    & 6.00975&0.9032258\\
    &-0.275\,532\,160827 &         &        &-0.261\,170\,846\,1206 &        &        \\
30.0&-0.257\,715\,275    & 8.543   &0.873979&-0.257\,278\,714\,7    & 8.5374&0.934896\\
    &-0.257\,715\,2848286&         &        &-0.257\,278\,715\,0364 &       &        \\
40.0&-0.254\,044\,456    &11.028   &0.904351&-0.254\,036\,841\,34   &11.02791&0.95097747\\
    &-0.254\,044\,4736052&         &        &-0.254\,036\,841\,4156 &        &        \\
50.0&-0.252\,534\,779    &13.522   &0.922873&-0.252\,534\,676\,43   &13.52162&0.96066311\\
    &-0.252\,534\,7813992&         &        &-0.252\,534\,676\,601  &        &         \\
\hline \hline
\end{tabular}
\end{center}
\end{table}

\subsection{Separation constant $A$}

In developed perturbation theory so as to estimate the accuracy of the
approximation \re{appr} for $X_0(\xi)$ and $Y_0(\eta)$, two
expressions, one for each variable, for the separation constant are
derived, $A_{n,\xi}\ $ and $ A_{n,\eta}$ (see Appendix and
Eqs. \re{xA_n} and \re{yA_n}). They are not independent: the condition
of consistency $A_{n, \xi}\ =\ A_{n,\eta}$ should be imposed. Table \ref{Aval}
presents the separation constant for all considered states. For each
$R$-value the first/second line correspond to $A_{\xi}$ /$A_{\eta}$
calculated with \re{xA_n} / \re{yA_n} compared to Madsen and Peek
\cite{Mar:1971} (third row). It turns out that as a result of
variational calculations the condition $A_{n,\xi}\ =\ A_{n,\eta}$ is
fulfilled automatically, up to $\sim 8$ s. d. which is in
agreement with those presented by Madsen and Peek
\cite{Mar:1971}. Hence, there is no need to impose the equality
condition. It is a reflection of the very high accuracy of the
approximation (17).

\begin{sidewaystable}
\caption{Separation parameters $A_{1,\xi}$ (first row), $A_{1,\eta}$ (second row)  for all the states considered of the H$_2^+$-ion compared to Madsen and Peek \cite{Mar:1971}(third row).
The results by Scott et al. \cite{Scott:2006} are in complete agreement with present calculations}
\label{Aval}
{\scriptsize
\begin{center}
\begin{tabular}{rllllllll}
\hline \hline
&$(0 0 0 +)$&$( 0 0 0 -)$&$( 0 0 1 +)$&$(0 0 1 -)$&$( 0 0 2 +)$&$(0 0 2 -)$&$( 1 0 0 +)$&$( 1 0 0 -)$\\
R[a.u.]&$1s\si_g$&$2p\si_u$&$2p\pi_u$&$3d\pi_g$&$3d\de_g$&$4f\de_u$&$2s\si_g$&$3p\si_u$\\
\hline
 1.0&  0.2499462430   & -1.8300104198  &  0.0476692616  & -3.9520464219   & 0.0157049965  &-5.9791583275  &  0.0711543055  & -1.9281072878  \\
    &  0.2499462409   & -1.8300104197  &  0.0476693150  & -3.9520464344   & 0.0157049889  &-5.9791583064  &  0.0711543140  & -1.9281072817  \\
    &  0.2499462406113& -1.830010419730&  0.047669315711& -3.952046434393 & 0.015704988875&-5.979158306119&  0.071154314127& -1.928107280448\\
 2.0&  0.8117295877   & -1.1868893947  &  0.1749484742  & -3.8048856116   & 0.0611354153  &-5.9165512457  &  0.2484661667  & -1.6917231809  \\
    &  0.8117295852   & -1.1868893929  &  0.1749484725  & -3.8048856050   & 0.0611354010  &-5.9165512311  &  0.2484661712  & -1.6917231733  \\
    &  0.8117295846248& -1.186889392359&  0.174948472433& -3.804885604702 & 0.061135400906&-5.916551230876&  0.248466171440& -1.691723172798\\
 4.0&  2.7995887561   &  1.5384644804  &  0.6001486772  & -3.1948053489   & 0.2270652065  &-5.6657454590  &  0.8535318015  & -0.7976034401  \\
    &  2.7995887582   &  1.5384644803  &  0.6001486748  & -3.1948053506   & 0.2270652107  &-5.6657454689  &  0.8535318003  & -0.7976034382  \\
    &  2.799588759471 &  1.538464480300&  0.600148674671& -3.194805350518 & 0.227065210827&-5.665745469006&  0.853531800197& -0.797603437898\\
 6.0&  6.4536037434   &  5.9279301781  &  1.2199716980  & -2.1786687874   & 0.4743694112  &-5.2501595578  &  1.8115068883  &  0.5663869192  \\
    &  6.4536037423   &  5.9279301759  &  1.2199717011  & -2.1786687836   & 0.4743694166  &-5.2501595612  &  1.8115068932  &  0.5663869192  \\
    &  6.453603742887 &  5.927930173726&  1.219971701568& -2.178668782566 & 0.474369416805&-5.250159561131&  1.811506894227&  0.566386919545\\
 8.0& 12.2261746132   & 12.0646853402  &  2.0537173294  & -0.7961022597   & 0.7914989890  &-4.6781903409  &  3.2069680505  &  2.3733521986  \\
    & 12.2261746118   & 12.0646853394  &  2.0537173246  & -0.7961022613   & 0.7914989805  &-4.6781903532  &  3.2069680527  &  2.3733521972  \\
    & 12.22617461542  & 12.06468533824 &  2.053717323829& -0.7961022613695& 0.791498980083&-4.678190353126&  3.206968053370&  2.373352197778\\
10.0& 20.1333096527   & 20.0921239053  &  3.1610270665  &  0.9355443423   & 1.1760019683  &-3.9601419353  &  5.1293596287  &  4.6288376336  \\
    & 20.1333042259   & 20.0921157054  &  3.1610270649  &  0.9355443394   & 1.1760019677  &-3.9601419604  &  5.1293596245  &  4.6288376291  \\
    & 20.13329317839  & 20.09209890008 &  3.161027064845&  0.9355443386850& 1.176001967652&-3.960141960690&  5.129359623687&  4.628837627894\\
20.0& 90.0528911866   & 90.0528775638  & 15.6431425753  & 15.4372141472   & 4.4202357771  & 1.6768434995  & 23.1467951638  & 23.1310108444  \\
    & 90.0528911837   & 90.0528775637  & 15.6431424784  & 15.4372141468   & 4.4202357567  & 1.6768434549  & 23.1467951625  & 23.1310108423  \\
    & 90.05289119141  & 90.05287756706 & 15.64314256883 & 15.43721414965  & 4.420235762270& 1.676843453846& 23.14679516399 & 23.13101084191 \\
30.0&210.0345966014   &210.0345966014  & 41.5927047072  & 41.5865009061   &11.8536327107  &11.1439910435  & 54.1918175098  & 54.1915412139  \\
    &210.0345965987   &210.0345965997  & 41.5927046648  & 41.5865009042   &11.8536321491  &11.1439910147  & 54.1918174666  & 54.1915412094  \\
    &210.0345965903   &210.0345965883  & 41.59270470411 & 41.58650090379  &11.85363268535 &11.14399101596 & 54.19181751174 & 54.19154120499 \\
40.0&380.0257071902   &380.0257071902  & 80.2475884726  & 80.2474668189   &25.5692520539  &25.4727279329  & 97.8369229167  & 97.8369191379  \\
    &380.0257071899   &380.0257071899  & 80.2475883011  & 80.2474668173   &25.5692515860  &25.4727279007  & 97.8369229125  & 97.8369191305  \\
    &380.0257071871   &380.0257071871  & 80.24758848264 & 80.24746682685  &25.56925202708 &25.47272792120 & 97.83692292343 & 97.83691912308 \\
50.0&600.0204520196   &600.0204516482  &131.4445904451  &131.4445885530   &45.2845813578  &45.2751100935  &154.0220957323  &154.0220957009  \\
    &600.0204519899   &600.0204516470  &131.4445904398  &131.4445885530   &45.2845807868  &45.2751100873  &154.0220957308  &154.0220956952  \\
    &600.0204516331   &600.0204516331  &131.4445904563  &131.4445885619   &45.28458134150 &45.27511009129 &154.0220957319  &154.0220956865  \\
\hline \hline
\end{tabular}
\end{center}}
\end{sidewaystable}

\section{Transitions}

Knowledge of wave functions with high local relative accuracy
$\lesssim 10^{-5} - 10^{-6}$ gives us a chance to calculate matrix
elements with controlled relative accuracy $\lesssim 10^{-5} -
10^{-6}$.  As a demonstration we calculate electric dipole, quadrupole and magnetic
dipole, E1, E2 and B1 Oscillator Strength as a function of interproton distance for
the permitted radiative transitions from excited states to the ground state
$1s\si_g\,(0,0,0,+)$ .

\subsection{E1 Oscillator Strength}

\begin{center}
\begin{figure}
\includegraphics[width=8cm]{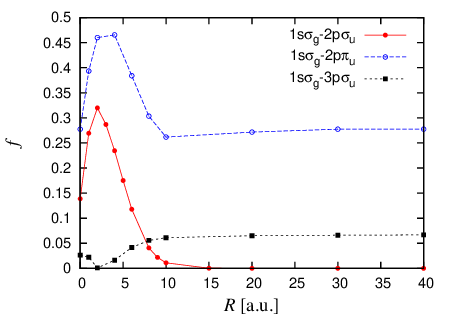}
\caption{$E1$ electric dipole oscillator strength $f$ between the some states and the ground state $1s\si_g$ (see text).}
\label{figDOS}
\end{figure}
\end{center}

Following Bates \cite{BDHS:1953,BDHS:1954}, with the energy given in
Rydbergs, the electric dipole oscillator strength from an lower electronic (initial)
state $\Psi_i$ to an  upper electronic (final) state $\Psi_f$ to a, is given by
\begin{equation}
\label{OSE1}
f_{i\rightarrow f}^{(E1)}(R) = \frac{1}{3}\,G\,(E_f(R)-E_i(R))\,{\bf S}^{(1)}_{if}\ ,
\end{equation}
where $G$ is the orbital degeneracy factor, ${\bf S}^{(1)}_{if}(R)$ is the square of the matrix element
\begin{displaymath}
 {\bf S}^{(1)}_{if}(R)=|\langle \Psi_i(R)| {\bf r} |\Psi_f(R)\rangle|^2\ ,
\end{displaymath}
and ${\bf r}$ is the vector of the electron position measured from the
interproton midpoint. The involved excited states for permitted
electric dipole transitions from the ground state $1s\si_g$ are three
states $2p\si_u$, $2p\pi_u$ and $3p\si_u$. In Table~\ref{Tose1} the
E1 oscillator strength is presented for two transitions:
$1s\si_g-2p\pi_u$ and $1s\si_g-3p\si_u$. The transition
$1s\si_g-2p\si_u$ was calculated and discussed in~\cite{Turbiner:2011}.
Here we present graphically these results on Fig.~\ref{figDOS}
in comparison with two other $E1$ electric dipole transitions. Certainly,
the $E1$ transition $1s\si_g-2p\pi_u$ is dominant for all $R$.
The orbital degeneracy factor is equal to $G=2$ for $f_{1s\si_g-2p\pi_u}$ and
$G=1$ for $f_{1s\si_g-3p\si_u}$. We assume this calculation should provide
at least 5 s.d. correctly.  As a result for all internuclear distances $R$
they coincide in 6 s.d. with Tsogbayar et al, \cite{ts:2010} for
$1s\si_g-2p\pi_u$ (with an exception at $R=1$\, a.u. where it deviates in one unit at the 6th
digit).  The present E1 oscillator strength $f_{1s\si_g-3p\si_u}$ is compared with
Bates et al~\cite{BDHS:1954} for $R=2,4$\,a.u., where the calculations were done in the past:
the agreement is within 2 s.d.  We confirm their striking
observation that the E1 oscillator strength increases in $\sim 20$
times coming from $R=2$\, a.u. to 4\,a.u. Furthermore, we observe a
dramatic dip in the E1 oscillator strength $f_{1s\si_g-3p\si_u}$ at $R
\sim R_{eq}$.  We do not have satisfactory physics arguments to
explain such a behavior.

\begin{table}[!thb]
  \caption{Electric dipole oscillator strength for transition
    $1s\si_g-2p\pi_u$ and $1s\si_g-3p\si_u$ vs $R$ compared to
    Tsogbayar et al~\cite{ts:2010}  and Bates et al~\cite{BDHS:1954}
    (rounded).}
\label{Tose1}
\begin{center}
\begin{tabular}{r|ll|ll}
\hline \hline
&\multicolumn{2}{c}{$f_{1s\sigma_g-2p\pi_u}$ $\times 10^{-1}$}
&\multicolumn{2}{c}{$f_{1s\sigma_g-3p\sigma_u}$ $\times 10^{-2}$}\\ \cline{2-5}
$R$[a.u.]&\ Present\ \ &\ \ \cite{ts:2010}\ \ &\ \ Present\ \ &\quad \ \cite{BDHS:1954}\\
\hline
  0.0 &\ 2.774 64 &             &\ 2.636 7 &\\
  1.0 &\ 3.934 37 &\ 3.934 381\ &\ 2.203 4 &\\
  2.0 &\ 4.601 87 &\ 4.601 870\ &\ 8.249  $\times 10^{-2}$\ &\ 8.24 $\times 10^{-2}$\\
  4.0 &\ 4.655 24 &\ 4.655 237\ &\ 1.614 4 &\ 1.61\\
  6.0 &\ 3.841 07 &\ 3.841 069\ &\ 4.146 0 &\\
  8.0 &\ 3.035 61 &\ 3.035 615\ &\ 5.567 8 &\\
10.0  &\ 2.617 50 &\ 2.617 505\ &\ 6.106 0 &\\
20.0  &\ 2.717 47 &\ 2.717 469\ &\ 6.503 4 &\\
30.0  &\ 2.774 38 &             &\ 6.610 8 &\\
40.0  &\ 2.775 81 &             &\ 6.673 4 &\\
50.0  &\ 2.775 50 &             &\ 6.715 3 &\\
\hline \hline
\end{tabular}
\end{center}
\end{table}

\subsection{B1 Oscillator Strength}

It is known that the magnetic dipole transitions are much smaller than
the electric dipole transition. The magnetic dipole B1 Oscillator
Strength, with the energy in Rydbergs, is given by
\begin{equation}
\label{OSB1}
f_{i\rightarrow f}^{(B1)}(R) = \frac{1}{3}\,(E_f(R)-E_i(R))|{\bf S}(R)|^2\ ,
\end{equation}
where ${\bf S}(R)$ is the matrix element
\begin{displaymath}
{\bf S}(R)=-\mu_B\langle \Psi_i(R)| {\bf L} |\Psi_f(R)\rangle\ ,
\end{displaymath}
${\bf L}$ is the angular momentum operator and $\mu_B$ is the Bohr
magneton. Between the states we consider at present article, there is
only one permitted magnetic dipole transition from the ground state to
$f_{1s\si_g-3d\pi_g}$. This B1 Oscillator strength is presented in
Table~\ref{TB1OS}. Comparison is made with previously known results by
Dalgarno et al.~\cite{DaMc:1953} at $R=2,4$~a.u. only with 3 s.d. We
confirm the striking qualitative observation made in \cite{DaMc:1953}
that the B1 oscillator strength increases in $\sim$10 times coming
from $R=2$~a.u. to 4~a.u. In general, it reflects extremely sharp
growth of the B1 Oscillator strength $f_{1s\si_g-3d\pi_g}$ at small
$R$: from $R=1$~a.u. to 2~a.u. it grows in $\sim 15$ times. In total,
from $R=1$\,a.u. to 4\,a.u. the B1 Oscillator strength
$f_{1s\si_g-3d\pi_g}$ increases in $\sim 200$ times!
It is related with the fact that at united atom limit, $R=0$, this
transition is prohibited but gets permitted at $R \neq 0$.

\begin{table}[h]
\caption{Magnetic dipole oscillator strength for transition $1s\si_g-3d\pi_g$ vs $R$ compared to
Dalgarno et al.~\cite{DaMc:1953}.}
\label{TB1OS}
\begin{center}
\begin{tabular}{rll}
\hline \hline
&\multicolumn{2}{c}{$f_{1s\si_g-3d\pi_g}$}\\ \cline{2-3}
$R$[a.u.]\ &\quad Present\ &\quad \cite{DaMc:1953}\\
\hline
 0.0\ &\quad 0.0          &\\
 1.0\ &\quad 1.050 61 E-08\ &\\
 2.0\ &\quad 1.666 18 E-07\ &\quad 1.67 E-07\\
 4.0\ &\quad 2.008 47 E-06\ &\quad 2.01 E-06\\
 6.0\ &\quad 6.251 64 E-06\ &\\
 8.0\ &\quad 1.129 24 E-05\ &\\
10.0\ &\quad 1.633 84 E-05\ &\\
20.0\ &\quad 5.260 00 E-05\ &\\
30.0\ &\quad 1.169 17 E-04\ &\\
40.0\ &\quad 2.078 47 E-04\ &\\
50.0\ &\quad 3.247 42 E-04\ &\\
\hline \hline
\end{tabular}
\end{center}
\end{table}
\subsection{E2 Oscillator Strength}

It is known that the electric quadrupole transitions are much smaller
than the electric dipole transition but comparable with magnetic
dipole transitions. For the first time we calculate electric
quadrupole transitions in $H_2^+$ molecular ion for transitions
${1s\si_g-3d\pi_g}$, ${1s\si_g-3d\de_g}$ and ${1s\si_g-2s\si_g}$.

The electric quadrupole E2 oscillator strength with the energy in
Rydbergs is given by
\begin{equation}
\label{OSE2}
f_{i\rightarrow f}^{(E2)}(R) = \frac{\alpha^2}{240}\,G\,(E_f(R)-E_i(R))^3 {\bf S}^{(2)}_{if}(R)\ ,
\end{equation}
where ${\bf S}^{(2)}_{if}(R)$ is the square of the matrix element of
the electric quadrupole moment and $\al$ is the fine structure
constant. The orbital degeneracy factor is $G=2$ for
$f_{1s\si_g-3d\pi_g}$ and $f_{1s\si_g-3d\de_g}$ and $G=1$ for
$f_{1s\si_g-2s\si_g}$. It is assumed this calculation should provide
at least 5 s.d. correctly. Results are presented in
Table~\ref{Tose2}. Comparing the electric dipole transition
$f_{1s\si_g-2p\pi_u}$, see Table \ref{Tose1} with the magnetic dipole
transition $f_{1s\si_g-3d\pi_g}$, see Table \ref{TB1OS}, and electric
quadrupole transition $f_{1s\si_g-3d\pi_g}$, see Table \ref{Tose2}
oscillator strengths, one can see that at $R=2$\, a.u. the E1 oscillator
strength is six orders of magnitude larger than E2 oscillator strength
and seven order of magnitude larger than B1. We have to pay attention
to exceptionally fast growth of the E2 oscillator strength
${1s\si_g-2s\si_g}$ in domain $R=1. - 4.$\,a.u. in $\sim 200$ times!
It is related with the fact that at united atom limit, $R=0$, this
transition is prohibited but gets permitted at $R \neq 0$.

\begin{table}[!htb]
\caption{Quadrupole oscillator strength $f$ for transitions ${1s\si_g-3d\pi_g}$, ${1s\si_g-3d\de_g}$ and ${1s\si_g-2s\si_g}$ {\it vs} $R$.}
\label{Tose2}
\begin{center}
\begin{tabular}{r|l|l|ll}
\hline \hline
$R$[a.u.]\quad &\quad $f_{1s\si_g-3d\pi_g}$\quad &\quad $f_{1s\si_g-3d\de_g}$ \quad &\quad
$f_{1s\si_g-2s\si_g}$\\
\hline
 0.0 \quad &\quad 3.744 24 E-07 \quad &\quad 3.744 24 E-07 \quad &\quad 0.0  \\
 1.0 \quad &\quad 1.500 69 E-06 \quad &\quad 1.240 33 E-06 \quad &\quad 1.386 51 E-09\\
 2.0 \quad &\quad 2.608 64 E-06 \quad &\quad 1.557 36 E-06 \quad &\quad 1.378 38 E-08\\
 4.0 \quad &\quad 4.539 82 E-06 \quad &\quad 1.436 91 E-06 \quad &\quad 1.372 68 E-07\\
 6.0 \quad &\quad 6.122 02 E-06 \quad &\quad 9.655 52 E-07 \quad &\quad 5.240 45 E-07\\
 8.0 \quad &\quad 7.884 70 E-06 \quad &\quad 5.901 76 E-07 \quad &\quad 1.222 38 E-06\\
10.0 \quad &\quad 1.010 48 E-05 \quad &\quad 3.817 37 E-07 \quad &\quad 2.179 70 E-06\\
20.0 \quad &\quad 3.114 40 E-05 \quad &\quad 1.558 01 E-07 \quad &\quad 9.918 09 E-06\\
30.0 \quad &\quad 6.984 15 E-05 \quad &\quad 1.735 37 E-07 \quad &\quad 2.253 65 E-05\\
40.0 \quad &\quad 1.244 75 E-04 \quad &\quad 1.864 75 E-07 \quad &\quad 4.027 41 E-05\\
50.0 \quad &\quad 1.946 58 E-04 \quad &\quad 1.879 85 E-07 \quad &\quad 6.317 65 E-05\\
\hline \hline
\end{tabular}
\end{center}
\end{table}
\section{H$_2^+$ molecular ion in the united atomic ion He$^+$ limit}

When for H$_2^+$ molecular ion the internuclear distance tends to
zero, $R \rightarrow 0$, we arrive at one-electron atomic system with
nuclear charge $Z=2$, {\it i.e.} the He$^+$ ion. In practice, at $R
\rightarrow 0$ we have
\begin{eqnarray}
\label{ellx}
  \lim_{R\rightarrow 0}R\,\xi\ =\ & 2 r\ ,\quad 0\leq r\leq \infty \ ,\\
  \lim_{R\rightarrow 0}\eta  \ =\ & \cos{\theta}\  ,\quad 0 \leq \theta \leq \pi \ ,\\
  \lim_{R\rightarrow 0}\phi  \ =\ & \phi \  ,\quad 0 \leq \phi \leq 2\pi \ ,
\end{eqnarray}
where $(r,\theta,\phi)$ are the spherical coordinates. However,
although in this limit the parameter $p \rightarrow 0$, the ratio
\begin{equation}
\lim_{R\rightarrow 0} \frac{R}{p} = \frac{2}{\sqrt{-E}}= \frac{2{\mathtt n}}{\mathcal{Z}}\Bigg|_{\mathcal{Z}=2}={\mathtt n}\,,
\end{equation}
(cf. \re{p}), takes a finite value; here $E=-\mathcal{Z}^2/
\mathtt{n}^2$ is the total energy of the hydrogen-like atom of
$\mathcal{Z}$-charge ($\mathcal{Z}=2$) with principal quantum number
$\mathtt{n}$.  Now taking the variational parameters $\al \rightarrow
0$, $\gamma\sim$ {\it const}, $a_{1} \rightarrow 0$
$b_{2}=b_{3}\rightarrow0$, the limit of approximation \re{appr} at
$R\rightarrow 0$ (up to a normalization factor) is
\begin{equation}
\Psi^{(\pm)}_{n,m,\La;\mathtt{n}} 	\propto r^{\mathtt{n}-n-1}P_n(r)
  e^{-\frac{2}{\mathtt{n}}r} \sin^{\Lambda}{\tha}\,\,Q_m(\cos^2{\tha})
 \left[
   \begin{array}{c}
    1  \\
    \cos{\tha}
    \end{array}
 \right] e^{\pm i \La \phi} \ .
\label{plim}
\end{equation}
This formulas realizes the correspondence between the states of the
molecular ion H$_2^+$ and ones of the atomic ion He$^+$. The examples
of this correspondence are displayed in Table~\ref{Rto0}. The first
column presents the molecular orbital $(n,m,\La,\pm)$ approximated
by~\re{appr}. Its united atom nomenclature is given in the second
column. In the limit $R \rightarrow 0$ this approximation takes the
form~\re{plim} (third column). Clearly, these functions coincide to
the exact wavefunctions of the atomic ion He$^+$ (up to normalization
factor), when the constants in the polynomial $P_n(r)$  or
$Q_m(\cos^2{\theta})$ (when present) take a certain values (see the
third column). Hence, the molecular orbital $(n,m,\La,\pm)$ in
approximation (\re{appr}) in the limit $R \rar 0$ corresponds to the
exact atomic orbital $\mathtt{(n,l,m)}$ with appropriate value of $l$,
as given in the fourth column of Table~\ref{Rto0}. In the opposite limit, $R
\rightarrow \infty$, the H$_2^+$  ion dissociates into a proton
plus a Hydrogen atom in the state with principal quantum number $N$ :
H$_2^+ \rightarrow$ $p$ + H- $\mbox{atom}[N]$, (shown in the last column).

\begin{table}[!thb]
  \caption{Correspondence between the molecular orbital $(n,m,\Lambda,\pm)$
    and the atomic orbital $\mathtt{(n,l,m)}$ in the limit $R \rightarrow 0$,
    here molecular approximation~\re{appr} takes the
    form~\re{plim}. In the limit $R \rightarrow \infty$, the ion
    H$_2^+$ dissociates into a proton plus a Hydrogen atom in the state
    with principal quantum number $N$ : H$_2^+ \rightarrow$ $p$ + H-
    $\mbox{atom}[N]$ (last column).}
\label{Rto0}
\begin{center}
\begin{tabular}{ll|lc|c}
\hline\hline
\multicolumn{2}{c}{Molecular Orbital}\vline&\multicolumn{2}{c}{Limit
  $R\rightarrow 0$}\vline&Limit $R\rightarrow \infty$\\
\hline
& & United Atom  &  Atomic Orbital&H$^+$ + H[N]\\
$(n,m,\La,\pm)$ & Designation &\ \re{plim}  & $\mathtt{(n,l,m)}$&N\\
\hline
$(0,0,0,+)$ & $1s\si_g$ &\ $e^{-2r}$  & $\mathtt{(1,0,0)}$&1\\
$(0,0,0,-)$ & $2p\si_u$ &\ $re^{-r}\cos{\theta}$  & $\mathtt{(2,1,0)}$&1\\
$(0,0,1,+)$ & $2p\pi_u$ &\ $re^{-r}\sin{\theta}e^{i\phi}$  & $\mathtt{(2,1,1)}$&2\\
$(0,0,1,-)$ & $3d\pi_g$ &\ $r^2e^{-\frac{2}{3}r}\sin{\theta}\cos{\theta}\,e^{i\phi}$
       & $\mathtt{(3,2,1)}$&2\\
$(0,0,2,+)$ & $3d\de_g$&$\ r^2e^{-\frac{2}{3}r}\sin^2{\theta}\,e^{2i\phi}$
       & $\mathtt{(3,2,2)}$&3\\
$(0,0,2,-)$ & $4f\de_u$ & $\ r^3e^{-\frac{1}{2}r}\sin^2{\theta}\cos{\theta}\,e^{2i\phi}$
       & $\mathtt{(4,3,2)}$&3\\
$(1,0,0,+)$ & $2s\si_g$ & $\ (r-2)e^{-2r}$ & $\mathtt{(2,0,0)}$&2\\
$(1,0,0,-)$ & $3p\si_u$ & $\ r(r-3)e^{-\frac{2}{3}r}\,\cos{\theta}$ & $\mathtt{(3,1,0)}$&2\\
$(0,1,0,+)$ & $3d\si_g$ & $\ r^2e^{-\frac{2}{3}r}(\cos^2{\theta}-1/3)$
       & $\mathtt{(3,2,0)}$&2\\
$(0,1,0,-)$ & $4f\si_u$ & $\ r^3e^{-\frac{1}{2}r}(\cos^2{\theta}-3/5)\,\cos{\theta}$
       & $\mathtt{(4,3,0)}$&2\\
\hline\hline
\end{tabular}
\end{center}
\end{table}


\section{The lowest states potential curves}

\subsection{Energy gap between $1s\si_g$ and $2p\si_u$ states}

The Born-Oppenheimer approximation leads to the concept of potential
curve, which is nothing but the total energy of the system H${}^+_2$
as a function of the internuclear distance $R$. Thus, the problem to
find a potential curve is reduced to finding spectra of electronic
Schr\"odinger equation (\ref{Sch}), where $R$ plays a role of
parameter. Since the potential in (\ref{Sch}) is a double-well
potential with degenerate minima, it is natural to study the energy
gap, which is the distance between two lowest eigenstates,
\begin{equation}
\label{gap}
      \De E\ =\ E_{2p\si_u}\ -\ E_{1s\si_g}\ .
\end{equation}
For small $R$ it was found~\cite{BB:1965,BS:1966,K:1983}
\begin{equation}
\label{DE-0}
  \De E\ =\ 3\ -\ \frac{27}{5}\,R^2\ +\ O\,(R^3)\ ,
\end{equation}
while at large $R$ \cite{OV:1964,DP:1968,Cizek:1986},
\begin{equation}
\label{DE-infty}
  \De E\ =\ 8 R e^{-R-1}\left(1 + \frac{1}{2R} -\frac{25}{8\, R^2} + \cdots\right)\ + O(e^{-2R}).
\end{equation}
It looks like the multi-instanton expansion where $R$ is the classical
action.

\begin{figure}[!thb]
\includegraphics[scale=1.0]{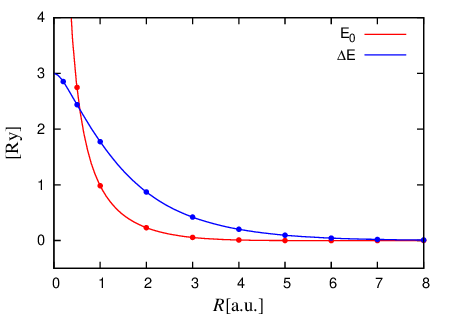}
\caption{$E_0$ and $\De E$ as defined by~\re{E0-sum} and~\re{gap},
  respectively. Calculated energies are marked by dots, the solid curves
  are the fits~\re{eRfit} and \re{f2}.}
\label{e0anddE}
\end{figure}

Now we take data for potential curves of the $1s\si_g$ and $2p\si_u$
states, see Tables \ref{Ten1ssg}, \ref{Ten2psu}, calculate the
difference $\De E$ and interpolate between small and large distances
using the Pad\'e approximation $e^{-R-1}\ \mbox{Pade}[N+1/N](R)$. In general,
$\De E$ is smooth, slow-changing curve with $R$, see below Fig.~\ref{e0anddE}.

\begin{itemize}

\item $e^{-R-1}\ \mbox{Pade}[4/3](R)$
\begin{equation}
\label{f1}
    \De E =e^{-R-1}\ \frac{3e+a_1 R+a_2 R^2+a_3 R^3+5 R^4}{1+\al_1 R+\al_2 R^2+\frac{5}{8}R^3}\ ,
\end{equation}
where a constraint
\begin{equation}
\begin{array}{l}
  \mbox{$\alpha_1= (a_1-3e)/(3e)$}\ ,\\
  \mbox{$\alpha_2=(a_3-5/2)/8$}\ ,\\
\end{array}
\end{equation}
is imposed.
After making fit with~(\ref{f1}), the 3 free parameters are found:
\begin{equation}
\label{parmsfits}
\begin{array}{l}
 a_1\ =\ 76.936\ ,\\
 a_2\ =\ 32.388\ ,\\
 a_3\ =\ -8.283\ .
\end{array}
\end{equation}
Approximant (\ref{f1}) reproduces correctly the $R^0$ and $R$ terms in
expansion (\ref{DE-0}) and the two terms in expansion
(\ref{DE-infty}). Eventually, this fit gives 2-3\, d.d. in average for $R \in [0,
40]$~a.u. which gets better larger $R$.
\item $e^{-R-1}\ \mbox{Pade}[8/7](R)$
\begin{equation}
\label{f2}
      \De E \ =\ e^{-R-1}\ \frac{3e+a_1 R+a_2 R^2+a_3 R^3+a_4 R^4+a_5 R^5+a_6 R^6
      +a_7 R^7+8 R^8} {1+\al_1 R+\al_2 R^2+b_3 R^3+b_4 R^4+b_5 R^5+\al_3 R^6+ R^7}\ ,
\end{equation}
where a constraint
\begin{equation}
\begin{array}{l}
   \mbox{$\alpha_1= (a_1-3e)/(3e)$}\ ,\\
   \mbox{$\alpha_2= (-a_1+a_2 +\frac{69e}{10})/(3e)$}\ ,\\
   \mbox{$\alpha_3=(a_7-4)/8$}\ ,
\end{array}
\end{equation}
is imposed.
After making fit with~(\ref{f2}),  the 10 free parameters become:
\begin{equation}
\label{parmsfits}
\begin{array}{ll}
a_1 = 605.5786\ ,\  &  a_6 =  410.7492\ , \\
a_2 = 1502.141\ ,\  &  a_7 = -80.17782\ , \\
a_3 = 2772.938\ ,\  &  b_3 =  316.7166\ , \\
a_4 = 762.9481\ ,\  &  b_4 = -154.1665\ , \\
a_5 =-757.1069\ ,\  &  b_5 =  59.71554\ .
\end{array}
\end{equation}
(\ref{f2}) reproduces correctly the $R^0$, $R^1$ and $R^2$ terms of
(\ref{DE-0}) and the two terms in (\ref{DE-infty}). This fit gives, in
general, 5-6\,d.d. at $R \in [0,40]$, and $\gtrsim$ 9\,d.d. at small $R \leq 1$\,a.u. and
up to 10\,d.d. for large $R \in [20,40]$\,a.u. \ (see for illustration
Fig.~\ref{e0anddE}).
\end{itemize}

\subsection{The ground state $1s\si_g$ and the first excited state $2p\si_u$}

For the lowest state $1s\si_g$, the behavior of the potential curve
$E_{1s\si_g}$ at the two asymptotic limits of small and large distances is
well known.  For $R\rightarrow 0$ the total energy is given
by~\cite{BB:1965,BS:1966,K:1983}
\begin{equation}
\label{R0-0}
    E_{1s\si_g}^{(0)}\ =\ \frac{2}{R}\ -\ 4\ +\ \frac{16}{3}\,R^2\ -\ \frac{32}{3}\,R^3\ +\ O(R^4 \log R) \ .
\end{equation}
Choosing the reference point for the energy at zero the behavior at
$R\rightarrow \infty$ reads~\cite{OV:1964,DP:1968,Cizek:1986}
\begin{eqnarray}
\label{vinfg}
  E_{1s\si_g}^{(\infty)} & = & -\frac{9}{2\, R^4}\ -\ \frac{15}{R^6}\ -\ \frac{213}{2\, R^7}\ +
  \cdots\\\non
  && - 4R\,e^{-R-1}\left[1+\frac{1}{2\, R}-\frac{25}{8\, R^2}-\frac{131}{48\, R^3}
  -\frac{3923}{384\, R^4}\ +\ \cdots\right]\ +\ O(e^{-2 R})\ ,
\end{eqnarray}
where the first sum represents perturbation theory, the second one is
a type of one-instanton contribution etc. As for the lowest state of
the negative parity $2p\si_u$ large and small $R$-distance expansions
are known as well,
\begin{equation}
\label{R0-1}
    E_{2p\si_u}^{(0)}\ =\ \frac{2}{R}\ -\ 1\ -\ \frac{1}{15}R^2\ +\ \ldots\ ,
\end{equation}
at $R \rar 0$,
while the behavior for $R\rightarrow \infty$ is given by
Eq.~\re{vinfg} with sign changed from minus to plus in front of the
exponential term $\sim e^{-R}$.

Let us consider the sum of potential curves for
$1s\si_g$ and $2p\si_u$ states,
\begin{equation}
\label{E0-sum}
     E_0\ \equiv \ \frac{E_{1s\si_g}+E_{2p\si_u}}{2}\ .
\end{equation}
Its corresponding expansions are
\begin{equation}
\label{R0-sum}
 E_0\ =\ \frac{2}{R}\ -\ \frac{5}{2}\ +\ \frac{79}{30}R^2\ +\ \ldots\ ,
\end{equation}
at $R \rar 0$ and
\begin{equation}
\label{Rinfty-sum}
 E_0\ =\ -\frac{9}{2\, R^4}\ -\ \frac{15}{R^6}\ -\ \frac{213}{2\, R^7}\ +\ \ldots\
 +\ O(e^{-2R})\ ,
\end{equation}
at $R \rar \infty$. Now we assume that two-instanton contribution,
$\sim e^{-2R}$ a large $R$, (and possible higher exponentially-small contributions)
can be neglected and construct the analytic approximation for $E_0$ which
mimics the two asymptotic limits (\ref{R0-sum}), (\ref{Rinfty-sum}) using
Pad\'e approximations $E_0(R)=\frac{1}{R}\ \mbox{Pade}[N/N+3](R)$ with a certain $N$.
Concrete fit was made for $N=5$, where the Pad\'e approximation is of the form
$E_0(R)=\frac{1}{R}\ \mbox{Pade}[5/8](R)$,
\begin{equation}
\label{eRfit}
   E_0 =\frac{2+ a_1R+a_2R^2 +a_3R^3 +a_4R^4 -9R^5}{R (1 +\al_1R
  +\al_2R^2+b_3R^3+b_4R^4+b_5R^5-\al_3R^6-\al_4R^7+2R^8)}\ ,
\end{equation}
with a certain constraints imposed,
\begin{equation}
\begin{array}{l}
   \mbox{$\al_1= (a_1+5/2)/2$}\ ,\\
   \mbox{$\al_2= (10a_1+8a_2+25)/16$}\ ,\\
   \mbox{$\al_3= 2(a_3+30)/9$}\ ,\\
   \mbox{$\al_5= 2a_4/9$}\ .
\end{array}
\end{equation}
After making the fit with~(\ref{eRfit}),  we arrive to concrete values of seven free parameters:
\begin{equation}
\label{parmsfits}
\begin{array}{ll}
a_1 = -24.019\ ,\  & b_3 =  72.243\ ,\\
a_2 =  602.64\ ,\  & b_4 =  2.4395\ ,\\
a_3 = -339.79\ ,\  & b_5 = -80.269\ ,\\
a_4 =  86.850\ .\  & \\
\end{array}
\end{equation}
It provides $\gtrsim$\,3-4\,d.d. for all studied domain $R \in [1,40]$.
Its free parameters are also in complete agreement in 5-6~s.d. with coefficients in
the terms $R^{-1}$, $R^{0}$ and $R^{1}$ of expansion at $R\rightarrow
0$, see \re{R0-sum} and $R^{-4}$, $R^{-5}$ and $R^{-6}$ at
$R\rightarrow \infty$, in the $1/R$-expansion (\ref{Rinfty-sum}),
see Fig.~\ref{e0anddE}.

\begin{figure}[!thb]
\includegraphics[scale=1.0]{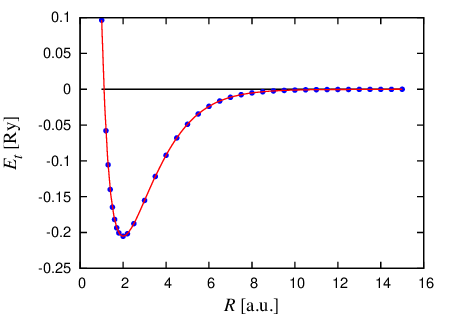}
\caption{Potential curve for the ground state $1s\si_g$: points are
  the calculated values and the solid curve is the fit~\re{pt1s}.}
\label{curve1ssgRy}
\end{figure}

In a consistent way, the potential curve for the ground state $1s\si_g$ can
be constructed from \re{eRfit} and \re{f2} by taking
\begin{equation}
\label{pt1s}
   E_{1s\sigma_g}\ =\ E_0 - \frac{1}{2}\De E\ .
\end{equation}
This expression reproduces 3-4\,d.d. when comparing with the exact
values, see Table \ref{Ten1ssg} and for illustration see
Fig.~\ref{curve1ssgRy}. The asymptotic expansions of Eq.~(\ref{pt1s})
are given by
\begin{eqnarray}
      &E_0^{}&=\frac{2}{R}\ -\ 4\ +\ 238.101R^2\ +\cdots \ ,\\
      &E_{\infty}&=-\frac{9}{2R^4}-\frac{15}{R^6}-\frac{24.035}{R^7}+\cdots
      -4Re^{-R-1}\left[ 1 + \frac{1}{2\,R}-\frac{3.11078}{R^2}\cdots\right] \ ,
\end{eqnarray}
which are in complete agreement with the first three terms at
$R\rightarrow 0$, and with the first three terms in the $1/R$
expansion and two terms in $1/R$ expansion of the pre-factor to
$e^{-R}$ for $R\rightarrow \infty$ (cf.~\re{R0-0} and \re{vinfg}).

\begin{figure}[!thb]
\includegraphics[scale=1.0]{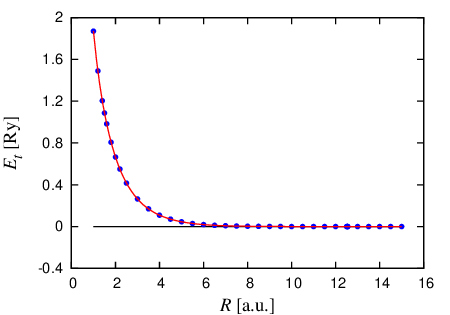}
\caption{Potential curve for the lowest state of negative parity $2p\si_u$: the
  calculated values are marked by dots and the solid curve represents the fit~\re{pt2p}.}
\label{curve2psuRy}
\end{figure}

Similarly, the potential curve for the excited state $2p\si_u$ is
restored from \re{eRfit} and \re{f2} by taking
\begin{equation}
\label{pt2p}
   E_{2p\sigma_u}\ =\ E_0 + \frac{1}{2}\De E\ .
\end{equation}
This expression also reproduces 3-4~d.d. when comparing with the exact
values, see Table \ref{Ten2psu} and for illustration
Fig.~\ref{curve2psuRy}. The asymptotic expansions of Eq.~(\ref{pt2p})
are given by
\begin{eqnarray}
     &E_0^{}&=\frac{2}{R}\ -\ 1\ +\ 232.701\,R^2\ +\cdots \ ,\\
     &E_{\infty}&=-\frac{9}{2R^4}-\frac{15}{R^6}-\frac{24.035}{R^7}+\cdots
      +4Re^{-R-1}\left[ 1 + \frac{1}{2\,R}-\frac{3.11078}{R^2}\cdots\right] \ ,
\end{eqnarray}
which are in complete agreement with the first three terms at
$R\rightarrow 0$ (cf.~\re{R0-1}), and three terms in $1/R$ expansion
and two terms in $1/R$ expansion of the pre-factor to $e^{-R}$ for
$R\rightarrow \infty$ (cf.~\re{vinfg}).

\section{Conclusions}

Summarizing we want to state that a simple uniform approximation of the eigenfunctions for the H$_2^+$ molecular ion is presented. It allows us to calculate any expectation value or matrix element with guaranteed accuracy. It manifests the approximate solution of the problem of spectra of the H$_2^+$ molecular ion. In a quite straightforward way similar approximations can be constructed for general two-center, one-electron system $(Z_a,Z_b,e)$, in particular, for $(\rm HeH)^{++}$ as well as for $(Z,Z,e)$. It will be done elsewhere.

The key element of the procedure is a straightforward interpolation between the WKB expansion at large distances and perturbation series at small distances for the phase of the wavefunction. Or, in other words, to find with high local accuracy an approximate solution for the corresponding eikonal equation. Separation of variables allowed us to solve this problem constructively. In the case of non-separability of variables the WKB expansion of a solution of the eikonal equation can not be constructed in unified way, since there is a strong dependence of the phase on the way to approach to infinity. However, a reasonable guess on the first growing terms of the WKB expansion seems sufficient to construct the interpolation between large and small distances which leads to highly accuracy results. This program is realized for the problem of the hydrogen atom in a constant magnetic field and will be published elsewhere.

In fact, with unusually high accuracy we are able to approximate the potential curves for the lowest states of positive, $1s\si_g$ and negative, $2p\si_u$ parity in the whole domain of the interproton distances, $R \geq 0$.
Eventually, the interproton interaction potential is described by a superposition of two suitable rational functions with $1/R$ and exponential in $R$ weight factors. It is different from the potentials used to approximate internuclear interaction in diatomic molecules (see \cite{Beckel:1980}, \cite{Sonnleitner:1980} and \cite{Warnicke:2015}, and references therein).
Analytic form of the approximation of the potential curve gives a chance to calculate
the corresponding vibrational states beyond harmonic approximation. Since long ago it
was known that at large $R$ these potential curves should contain exponentially-small
contributions, see e.g. \cite{OV:1964,DP:1968,Cizek:1986}, as a result of tunnelling
between two degenerate Coulomb wells. Energy difference between potential curves of
$1s\si_g$ and $2p\si_u$ states at $R \rar \infty$ should be exponentially small,
it can not be found in perturbation theory in $1/R$.
We are not aware about any calculations of this difference in instanton calculus.
Presence of the second term in generalized Pade approximation, $\De E \sim e^{-R}$,
see \re{pt1s} and \re{pt2p}, allows us to estimate for the first time the effect
of exponentially-small terms to a potential curve at finite $R$. This effect is
extremely small at large $R$ being $\sim 10^{-7}$ at $R=20$\,a.u. and giving contribution
to 11th s.d. and beyond for $R > 30$\,a.u. However, it becomes significant at $R < 20$\,a.u.,
see Fig. \ref{e0anddE}.


It is worth mentioning a curious fact that the problem (\ref{Sch})
after separation angular dependence possesses the hidden algebra
$sl(2)\oplus sl(2)$ \cite{Turbiner:2011}. The differential operator in $\xi$ and $\eta$
is in the universal enveloping algebra of $sl(2)\oplus sl(2)$ (see
e.g. \cite{Turbiner:1988}). The spin of the representation is
$-(\La+1)$ and $-\La-1+\frac{R}{p}$, respectively. For non-physical, (half)-integer,
positive values of $-\La$ and integer ratio $\frac{R}{p}$ the algebras $sl(2)$
appear in the finite-dimensional representation realized in action on
polynomials in $\xi, \eta$, respectively. It explains sometimes
observed a mysterious appearance of polynomial solutions for non-physical
values of $\La$ in the problem (\ref{Sch}).

\bigskip

\textit{\large \bf Acknowledgements}.

\bigskip

The research is supported in part by PAPIIT grant {\bf IN108815} and
CONACyT grant {\bf 166189} (Mexico). H.O.P. is grateful to
Universit{\'e} Libre de Bruxelles (Belgium) and Instituto de Ciencias
Nucleares, UNAM (Mexico) for a kind hospitality extended to him where
a certain stages of the present work were carried out. A.V.T. is grateful
to E Shuryak (Stony Brook) for the interest to work and encouragement.
A.V.T. gratefully acknowledges support from the Simons Center for Geometry and
Physics, Stony Brook University at which some of the research for this
paper was performed and where the paper was completed.

\bigskip
\section*{Appendix}
\label{ap1}
\bigskip

The easiest way to calculate a deviation of the approximation from the exact eigenfunction is to develop a perturbation theory in framework of the so-called {\it non-linearization procedure} \cite{Turbiner:1984}: for a chosen approximation $\psi_0$ a corresponding potential $V_0=\frac{\De \psi_0}{\psi_0}$ is found with $E_0=0$, for which $\psi_0$ is the exact eigensolution. Then the potential is written in the form $V=V_0 + \la V_1$, then it is looked for energy and the eigenfunction in the form of power series in the parameter $\la$, $E=\sum \la^n E_n$ and $\Psi=\Psi_0 \exp (-\sum \la^n \varphi_n)$, respectively. Eventually, $\la$ is placed equal to one.

Due to specifics of (1) because of the separation of variables the procedure can be developed for both functions $X$ and $Y$ (see (\ref{psi})) separately as well as for the separation parameter $A$, while keeping the energy $E$ fixed. It can be done for the system of equations (\ref{X_L}), (\ref{Y_L}). As a first step let us transform (\ref{X_L}), (\ref{Y_L}) into the Riccati form by introducing $X=fe^{-\varphi}$ and $Y=ge^{-\varrho}$, respectively,
\begin{equation}
\label{X_L-phi}
 (\xi^2-1)[f(x'- x^2)+2f'x-f''] + 2 (\La+1)\xi[fx-f'] =\
  [A - V(\xi)]f \ ,\quad x=\varphi'_{\xi}
\end{equation}
where the "potential" $V(\xi) = p^2\xi^2 - 2R \xi$, and
\begin{equation}
\label{Y_L-rho}
 (\eta^2-1)[g(y' - y^2)+2g'y-g''] + 2 (\La+1)\eta[gy-g'] =\
       [A - W(\eta)]g \ ,\quad y = \varrho'_{\eta}
\end{equation}
where the "potential" $W(\eta) = p^2\eta^2$.

Let us choose some $x_0(\xi)=\varphi_0'(\xi)$, then substitute it to the l.h.s. of (\ref{X_L-phi}) and call the result as unperturbed "potential" $V_0(\xi)$ putting without loss of generality $A_0=0$. The difference between the original $V(\xi)$ and generated $V_0(\xi)$ is the perturbation, $V_1(\xi)=V(\xi)-V_0(\xi)$. For a sake of convenience we can insert a parameter $\la$ in front of $V_1$ and develop the perturbation theory in powers of it. The perturbation theory is also developed for node states where a node position is also looked for the form of power expansion in $\la$.
\begin{equation}
\label{PTx}
    x=\sum \la^n x_n\ , \ f=\sum \la^n f_{n,\xi}\ , \ A=\sum \la^n A_{n,\xi}\ .
\end{equation}
The equation for $n$th correction has a form,
\begin{equation}
\label{x_n}
   \left\{(\xi^2-1)^{\La+1}X_0^2\left[x_n-\left(\frac{f_{n,\xi}}{f_{0,\xi}}\right)'\right]\right\}'
    =(\xi^2-1)^{\La}X_0^2[A_{n,\xi} - Q_n],
\end{equation}
where $Q_1=V_1$ and
\begin{eqnarray}
Q_n &=& -(\xi^2-1)\sum_{i=1}^{n-1}x_ix_{n-i}\nonumber\\
    & & -\frac{1}{f_{0,\xi}}\left[\sum_{k=1}^{n-1}f_{k,\xi}\left((\xi^2-1)\sum_{i=0}^{n-k}x_ix_{n-k-i}-\frac{((\xi^2-1)^{\La+1}x_{n-k})'}{(\xi^2-1)^{\La}} +A_{n-k,\xi}-V_{n-k}\right)\right.\nonumber\\
    & &	-\left.2(\xi^2-1)\sum_{k=1}^{n-1}x_kf'_{n-k,\xi}\right],
\end{eqnarray}
for $n>1$. Integrating~\re{x_n} we obtain
\begin{equation}
\label{x_n-solu}
x_n\ =\left(\frac{f_{n,\xi}}{f_{0,\xi}}\right)'+\frac{1}{(\xi^2-1)^{\La+1}X_0}\int_1^{\xi} (A_{n,\xi} - Q_n) (\xi^2-1)^{\La}X_0^2 \,d\xi\ ,
\end{equation}
where $f_{n,\xi}$  and $A_{n,\xi}$ are obtained in the same way. These are
\begin{equation}
\label{xf_n}
  f_{n,\xi}(\xi_0)\ =\frac{1}{(\xi_0^2-1)^{\La+1}e^{-2 \varphi_0}f'_{0,\xi}(\xi_0)} \int_1^{\xi_0}(A_{n,\xi} - Q_n) (\xi^2-1)^{\La}X_0^2\,d\xi,
\end{equation}
and
\begin{equation}
\label{xA_n}
    A_{n,\xi}\ =\ \frac{\int_1^{\infty} Q_n (\xi^2-1)^{\La}X_0^2d\xi}
    {\int_1^{\infty} (\xi^2-1)^{\La}X_0^2\,d\xi}.
\end{equation}
In a similar way by choosing $y_0(\eta)=\varrho_0'(\eta)$, building the unperturbed "potential" $W_0(\eta)$ and putting $A_0=0$ as zero approximation one can develop perturbation theory in the equation (\ref{Y_L-rho})
\begin{equation}
\label{PTy}
   y=\sum \la^n y_n\ , \ g=\sum \la^n g_{n,\eta}\, \ A=\sum \la^n A_{n,\eta}\ .
\end{equation}
The equation for $n$th correction has a form similar to (\ref{x_n}),
\begin{equation}
\label{y_n}
  \left\{(\eta^2-1)^{\La+1}Y_0^2\left[y_n-\left(\frac{g_{n,\eta}}{g_{0,\eta}}\right)'\right]\right\}'
   =(\eta^2-1)^{\La}Y_0^2[A_{n,\eta} - Q_n],
\end{equation}
where $Q_1=W_1$ and
\begin{eqnarray}
Q_n &=& -(\eta^2-1)\sum_{i=1}^{n-1}y_iy_{n-i}\nonumber\\
    & & -\frac{1}{g_{0,\eta}}\left[\sum_{k=1}^{n-1}g_{k,\eta}\left((\eta^2-1)\sum_{i=0}^{n-k}y_iy_{n-k-i}-\frac{((\eta^2-1)^{\La+1}y_{n-k})'}{(\eta^2-1)^{\La}} +A_{n-k,\eta}-V_{n-k}\right)\right.\nonumber\\
    & &	-\left.2(\eta^2-1)\sum_{k=1}^{n-1}y_{k}g'_{n-k,\eta}\right],
\end{eqnarray}
for $n>1$. Its solution is given by (cf.(\ref{x_n-solu}))
\begin{equation}
\label{y_n-solu}
y_n\ =\left(\frac{g_{n,\eta}}{g_{0,\eta}}\right)'+\frac{1}{(\eta^2-1)^{\La+1}Y_0}\int_{-1}^{\eta} (A_{n,\eta} - Q_n) (\eta^2-1)^{\La}Y_0^2 \,d\eta\ ,
\end{equation}
where $g_{n,\eta}$  and $A_{n,\eta}$ are obtained in the same way. These are (cf.(\ref{xf_n}) and \re{xA_n})
\begin{equation}
\label{yg_n}
  g_{n,\eta}(\eta_0)\ =\frac{1}{(\eta_0^2-1)^{\La+1}e^{-2 \varrho_0}g'_{0,\eta}(\eta_0)} \int_1^{\eta_0}(A_{n,\eta} - Q_n) (\eta^2-1)^{\La}Y_0^2\,d\eta,
\end{equation}
and
\begin{equation}
\label{yA_n}
    A_{n,\eta}\ =\ \frac{\int_{-1}^{1} Q_n (\eta^2-1)^{\La}Y_0^2d\eta}
    {\int_{-1}^{1} (\eta^2-1)^{\La}Y_0^2\,d\eta}.
\end{equation}

 In order to realize this perturbation theory a condition of consistency should be imposed
\begin{equation}
\label{An}
    A_{n,\xi}\ =\ A_{n,\eta}\ .
\end{equation}
This condition allows us to find the parameter $p$ and, hence, the energy $E'$ and $E$ (see (\ref{p})).

Sufficient condition for such a perturbation theory to be convergent is to require a perturbation "potential" to be bounded,
\begin{equation}
\label{PerPot}
    |V_1(\xi)| \leq C_{\xi}\ ,\ |W_1(\eta)| \leq C_{\eta}\ ,
\end{equation}
where $C_{\xi}, C_{\eta}$ are constants. Obviously, that the rate of convergence gets faster with smaller values of $C_{\xi}, C_{\eta}$. It is evident that the perturbations $V_1(\xi)$ and $W_1(\eta)$ get bounded if $\varphi_0(\xi)$ and $\varrho_0(\eta)$ are smooth functions vanishing at the origin but reproduce exactly the growing terms at $|\xi|, |\eta|$ tending to infinity in (\ref{X-inf}), (\ref{Y-inf}), respectively.

%
%

\end{document}